\documentclass[draft]{agujournal}
\draftfalse

\journalname{JGR-Space Physics}

\begin{document}

%
%


\title{How accurately can we measure the reconnection rate $E_M$ for the MMS diffusion region event of 2017-07-11?}

%
%




\authors{K. J. Genestreti\affil{1}\thanks{Currently at Space Science Center, University of New Hampshire, Durham, New Hampshire, USA},
              T. K. M. Nakamura\affil{1},
              R. Nakamura\affil{1},
              R. E. Denton\affil{2},
              R. B. Torbert\affil{3,4},
              J. L. Burch\affil{4},
              F. Plaschke\affil{1},
              S. A. Fuselier\affil{4,5},
              R. E. Ergun\affil{6},
              B. L. Giles\affil{7},
              C. T. Russell\affil{8}}


\affiliation{1}{Space Research Institute, Austrian Academy of Sciences, Graz, Austria}
\affiliation{2}{Department of Physics and Astronomy, Dartmouth College, Hanover, New Hampshire, USA}
\affiliation{3}{Space Science Center, University of New Hampshire, Durham, New Hampshire, USA}
\affiliation{4}{Space Science and Engineering Division, Southwest Research Institute, San Antonio, Texas, USA}
\affiliation{5}{Department of Physics and Astronomy, University of Texas at San Antonio, San Antonio, Texas, USA}
\affiliation{6}{Laboratory of Atmospheric and Space Sciences, University of Colorado Boulder, Boulder, Colorado, USA}
\affiliation{7}{Heliophysics Science Division, NASA Goddard Space Flight Center, Greenbelt, Maryland, USA}
\affiliation{8}{Institute of Geophysics and Planetary Physics, University of California Los Angeles, Los Angeles, CA, USA}





\correspondingauthor{Kevin J. Genestreti}{kevin.genestreti@oeaw.ac.at}




\begin{keypoints}
\item The reconnection rate $E_M$ is estimated for one event using several techniques to find an $M$ direction.
\item The error bars in $E_M$ and the $LMN$ coordinate directions are estimated from virtual data.
\item {The reconnection rate is likely $E_M$=3.2 mV/m $\pm$ 0.6 mV/m, which corresponds to a normalized rate of 0.18$\pm$0.035.}
\end{keypoints}

%
%


\begin{abstract}
We investigate the accuracy with which the reconnection electric field $E_M$ can be determined from in-situ plasma data. We study the magnetotail electron diffusion region observed by NASA's Magnetospheric Multiscale (MMS) on 2017-07-11 at 22:34 UT and focus on the very large errors in $E_M$ that result from errors in an $LMN$ boundary-normal coordinate system. We determine several $LMN$ coordinates for this MMS event using several different methods. We use these $M$ axes to estimate $E_M$. We find some consensus that the reconnection rate was roughly $E_M$=3.2 mV/m $\pm$ 0.06 mV/m, which corresponds to a normalized reconnection rate of $0.18\pm0.035$. Minimum variance analysis of the electron velocity (MVA-$v_e$), MVA of $E$, minimization of Faraday residue, and {{an adjusted}} version of the maximum directional derivative of the magnetic field (MDD-$B$) technique all produce {reasonably} similar coordinate axes. We use virtual MMS data from a particle-in-cell simulation of this event to estimate the errors in the coordinate axes and reconnection rate associated with MVA-$v_e$ and MDD-$B$. The $L$ and $M$ directions are most reliably determined by MVA-$v_e$ when the spacecraft observes a clear electron jet reversal. When the magnetic field data has errors as small as 0.5\% of the background field strength, the $M$ direction obtained by MDD-$B$ technique may be off by as much as 35$^\circ$. The normal direction is most accurately obtained by MDD-$B$. Overall, we find that these techniques were able to identify $E_M$ from the virtual data within error bars $\geq$20\%.
\end{abstract}

%
%

%


%
%
%
%

\section{Introduction}

\subsection{Calculating the reconnection rate from in-situ plasma data}

In-situ measurements of the normalized reconnection rate $\mathcal{R}$ have been made at the Earth's magnetopause \cite{Mozer.2002,Fuselier.2005,Chen.2017}, its magnetotail \cite{Wygant.2005,Xiao.2007}, and its magnetosheath \cite{Phan.2007}, in the magnetospheres of other planets such as Mercury \cite{Slavin.2009} and Saturn \cite{Arridge.2016}, in the solar wind \cite{Phan.2006}, and in laboratory experiments \cite{Egedal.2007}. Calculating $\mathcal{R}$ as the rate of change of magnetic connectivity is not possible in practice with in-situ space plasma observations, so proxies are typically used that are either directly or with few assumptions equivalent to $\mathcal{R}$. 

In the studies mentioned in the previous paragraph, $\mathcal{R}$ was defined by either (1) the upstream inflow speed normalized by the downstream outflow speed $\mathcal{R}=V_{in}/V_{out}=v_{Nb}/V_{Aib}$, {where $v_{Nb}$ is the inflow speed and $V_{Aib}$ is the ion Alfven speed in the inflow region}, (2) the component of the magnetic field normal to the current sheet normalized by the reconnecting magnetic field strength $\mathcal{R}=B_N/B_{b}$, {where $B_b$ is the strength of the reconnecting component of the magnetic field in the inflow region}, (3) the normalized tangential reconnection electric field $E_M/E_b=E_M/V_{Aib}B_{b}$, or (4) the angle of the {ion} outflow fan. Here, $\hat{M}$ is the direction of the reconnecting current sheet, $\hat{N}$ is the current sheet normal, and $(\pm)\hat{L}$ is the direction of the reconnecting magnetic fields. While the canonical fast reconnection rate is $\mathcal{R}=0.1$, the exact value of $\mathcal{R}$ may depend on the magnetic shear angle \cite{MozerandRetino.2007,FuselierandLewis.2011}, the mass density of minor ion species \cite{Wang.2014,Liu.2015}, the presence of external driving forces \cite{Nakamura.2017}, turbulence and anomalous resistivity effects \cite{Che.2017}, etc.  

\subsection{Effect of errors in the measured coordinate system}

The inflow speed $v_{Nb}$, the normal magnetic field $B_N$, and the tangential electric field $E_M$ are typically the smallest components of their associated vectors. Very large errors in $\mathcal{R}$ can result from, for example, $E_M$ being evaluated inaccurately as $E_M^\ast=\vec{E}\cdot\hat{M}^\ast$, where some measured axis $\hat{M}^\ast$ has, by error, a finite projection onto the current sheet normal such that $\hat{M}^\ast=\cos\theta_{NM\ast}\hat{M}+\sin\theta_{NM\ast}\hat{N}$ and, for sufficiently small $\theta_{NM\ast}$, $\mathcal{R}^\ast\approx (E_M+\theta_{NM\ast} E_N)/E_b$. Given that the normal electric field can be tens of times larger than the reconnection electric field \cite{MozerandRetino.2007,Shay.2016,Chen.2017,Torbert.2018}, a relatively small error of $\theta_{NM\ast}\sim5^\circ$ could create error bars for $\mathcal{R}$ of $\sim$100\%. (Note that throughout the rest of the manuscript, the asterisk is used to denote coordinate axes or quantities that are known to be inaccurate).

\subsection{Goals of this study}

Here, we investigate the accuracy with which we find the normalized reconnection electric field $\mathcal{R}=E_M/E_b$ for a reconnection electron diffusion region (EDR) event observed in the Earth's magnetotail by MMS on 2017-07-11 at $\sim$22:34 UT \cite{Torbert.2018} [\textit{Nakamura et al.}, submitted]. We focus on the errors in $\mathcal{R}$ that result from inaccuracies in the coordinate system rather than other sources of error that may arise from inaccuracies in the measured electric field, inaccurate determination of the upstream normalization parameter $E_b$, etc (see discussion in section 5). While many other techniques exist for estimating $\mathcal{R}$ (see for instance our companion paper, \textit{Nakamura et al.} [submitted], hereafter N18), these techniques may have their own unique sources of error that are largely beyond the scope of this study.

In the next sections we discuss the MMS data used in this study (section 2.1), the observations of MMS during the 2017-07-11 EDR event (section 2.2), the set-up of the PIC simulation of N18 (section 2.3), and the virtual probe data from the simulation of N18 (section 2.4). In section 3 we find several $LMN$ coordinate systems and reconnection rates from MMS data. In section 4 we apply some of the same analysis techniques to the virtual probe data, where $L$, $M$, $N$, and $\mathcal{R}$ are known and the errors associated with our methods for finding them can be calculated directly. Finally, in section 5, we summarize and discuss our findings.

\section{Overview of MMS data, 2017-07-11 reconnection event, and PIC simulation}

\subsection{MMS data}

The DC magnetic field data is provided by the fluxgate magnetometers (FGM) at 128 vectors-per-second during high-time-resolution burst mode and nominally at 8 vectors-per-second \cite{Russell.2016}. The spin-plane components ($\sim B_X$ and $\sim B_Y$) of the magnetic field are calibrated to a high degree of accuracy by removing spin-tone oscillations in a de-spun coordinate system. The spin-axis magnetic field component is cross-calibrated with data from the electron drift instrument \cite{Torbert.2016c}. The stated accuracy of the DC magnetic field is $\pm0.1$ nT. Using data from two quiet magnetotail periods before (22:10--22:20 UT) and after (22:39-22:51 UT) the EDR interval, we found average inter-probe differences that were of the order $\pm0.05$ nT for the spin-axis components, while the absolute differences between the spin-plane components were much smaller on average ($\sim$0.001 nT) but had small residual spin-tones with amplitudes of $\sim$0.02 nT. 

The coupled AC-DC electric field data is provided by the electric field double probes instruments at 8196 vectors-per-second during burst mode and at 32 vectors-per-second during fast survey mode \cite{Lindqvist.2016,Ergun.2016}. We use the level 3 version of the electric field data{, which were determined for this event by \cite{Torbert.2018} by cross-calibrating $-\vec{v}_e\times\vec{B}$ and $\vec{E}$ to remove offsets in the perpendicular components of $\vec{E}$ (c.f. \textit{Wang et al.} (2017))}. The nominal uncertainty in the perpendicular electric field is expected to be $\sim$0.5 mV/m \cite{Torbert.2016c}. 

High time resolution plasma ion and electron moments are obtained by the fast plasma investigation (FPI) suite of sensors \cite{Pollock.2016}. In burst mode, 3-d electron (ion) distribution functions and moments are measurement once every 30 (150) ms. In regions with very sparse plasma, portions of phase space are under-sampled, such that the number of counts per (energy-angle-angle) pixel are comparable to the Poisson uncertainty. This leads to noise in the plasma moments. Other issues related to the data quality of the ion measurements taken during this event are discussed in \cite{Torbert.2018}.

\subsection{Overview of the 11 July 2017 EDR event}

\begin{figure*}
\noindent\includegraphics[width=39pc]{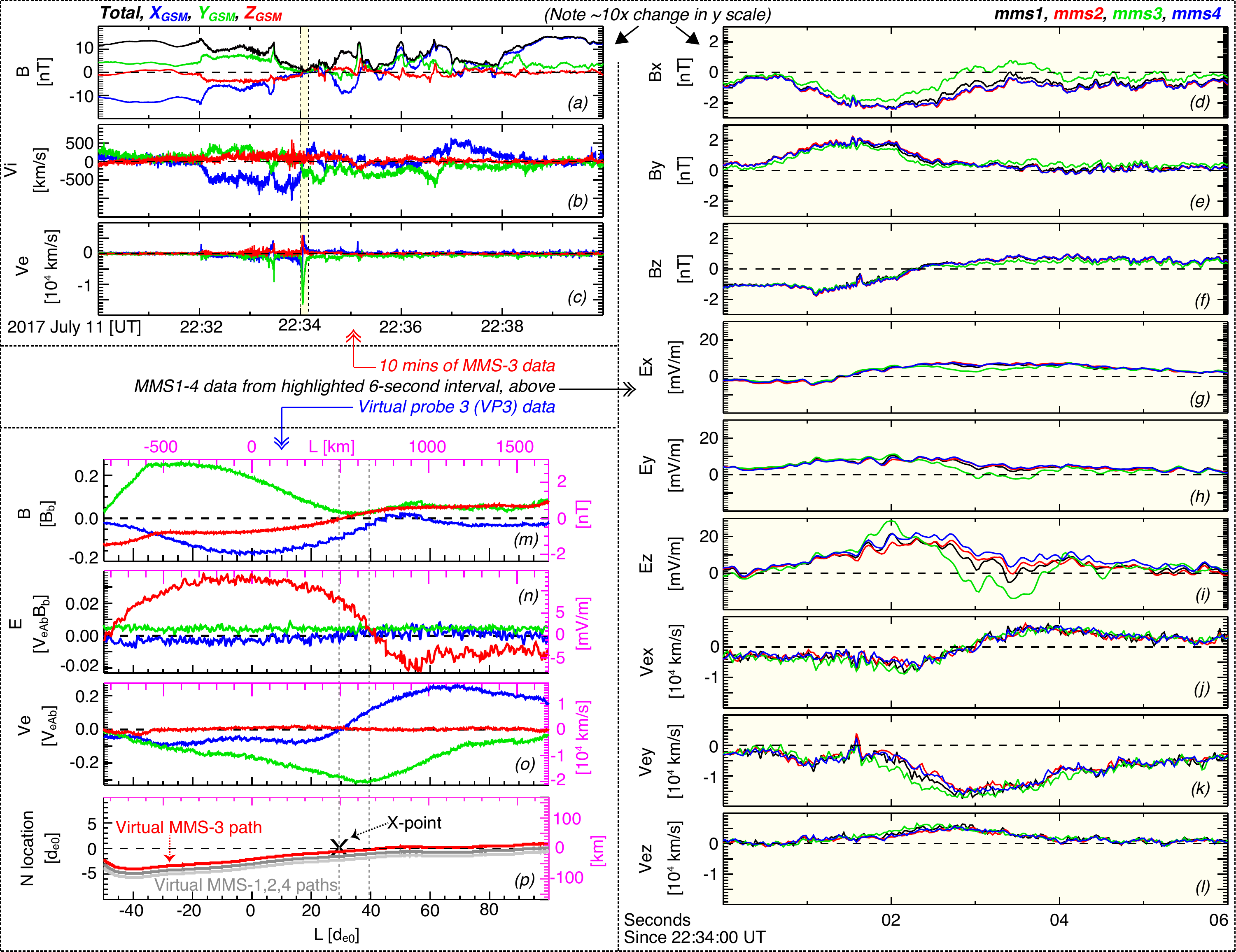}
\caption{(a) The magnetic field $\vec{B}$, (b) ion bulk velocity $\vec{v}_i$, and (c) electron bulk velocity $\vec{v}_e$ from MMS-3 from the roughly 10-minute current sheet crossing. The three components of (d)-(f) $\vec{B}$, (g)-(i) the electric field $\vec{E}$, and (j)-(l) $\vec{v}_e$ for the roughly 6-second EDR encounter, which is also indicated in the highlighted region in (a)-(c). (m) $\vec{B}$, (n) $\vec{E}$, (o) $\vec{B}$, and (p) the $N-L$ locations of the orbits of virtual MMS-3 (red), virtual MMS-1 (dark grey), and virtual MMS-2 and 4 (light grey). The vertical black and magenta-colored axes give the quantities in normalized and unnormalized units, respectively. MMS data are shown in GSM coordinates.}
\label{overview}
\end{figure*}

On 11 July 2017 at $\sim$22:34 UT, MMS observed an electron diffusion region (EDR) in the central magnetotail current sheet. The average inter-probe separation was approximately $\sim$17 km, which is approximately half of the asymptotic electron inertial length $d_{eb}\approx30$ km, and the formation was a near regular tetrahedron (tetrahedron quality factor of 0.957 \cite{Fuselier.2016}, see Figure \ref{tetrahedrons}a-b). The spacecraft was 22 earth radii ($R_E$) geocentric distance, 4 $R_E$ duskward of midnight, and less than 50 km (<0.007 $R_E$) away from the empirical-model-predicted neutral sheet location \cite{Fairfield.1980}, which was the most probable region for MMS to observe the diffusion region during the first magnetotail survey phase \cite{Genestreti.2014,Fuselier.2016}. An overview of the data from this event is given in Figure \ref{overview}, where data from the $\sim$10-minute current sheet crossing is in \ref{overview}a-c, data from the $\sim$6-second EDR crossing is in \ref{overview}d-l, and \ref{overview}m-p show the virtual MMS-3 data over a range comparable to \ref{overview}d-l. The virtual data will be discussed in section 2.4. Overall, during the 10-minute period shown in Figure \ref{overview}a-c, the spacecraft moved from the southern to the northern hemisphere {(see the negative-to-positive reversal in $B_X$ in Figure \ref{overview}a)} while crossing from the tailward to the earthward-pointing reconnection exhausts {(see the multiple bipolar variations in $B_Z$ with associated $|B|$ enhancements, e.g. at 22:36:00 UT and 22:36:40 UT, in Figure \ref{overview}b)}. Several ion-scale flux ropes are observed between the prolonged interval of tailward ion jetting and the prolonged interval of earthward jetting {(see the negative-to-positive reversal in $v_{iX}$ in Figure \ref{overview}a)}, some with intense electric fields, intense currents, and non-ideal energy conversion $\vec{J}\cdot\vec{E}'\neq0$ \cite{Zenitani.2011} {(not pictured)}. Primarily two of the quadrupolar Hall magnetic field ($\approx B_Y(X,Z)$) regions are observed, as well as both regions of the bipolar normal field ($\approx B_Z(X)$) and reconnecting field ($\approx B_X(Z)$). In the downstream separatrix (near $\sim$22:33:20 UT), intense parallel electron currents are observed along with intense electric fields and electron heating. In summary, the magnetic field observed during the 10-minute crossing is not that of a uniform 1-d current sheet, which is a common assumption of many techniques for finding $LMN$ coordinates. 

Parameters describing the initial state of the plasma sheet and lobes were detailed in N18. The plasma sheet ion and electron densities and temperatures were selected in the interval between 22:32-22:33 UT. The plasma sheet density was determined to be $n_0 \approx 0.08-0.1$ cm$^{-3}$ and the ion temperature was $T_{i0}\approx 4000-5000$ eV. The lobe densities and temperatures were determined near the EDR interval and near 22:33:30 during two brief excursions outside the reconnection separatrices. These lobe values were $T_{ib}\approx1000-2000$ eV and $n_b\approx$0.03 cm$^{-3}$. The electron temperatures were roughly one third of the ion temperatures. The lobe magnetic field was roughly $B_b\approx10-12$ nT. The variability and uncertainty in these parameters indicate that $E_b=V_{Aib}B_b$, the parameter that normalizes the reconnection rate $\mathcal{R}=E_M/E_b$, may not have been determined perfectly accurately. However, the best estimate for this parameter is $E_b$=18.12 mV/m, using the average values for the upstream Alfven speed $V_{Aib}$ and lobe magnetic field. The combined pressures derived from these lobe values are roughly in balance with the combined pressures determined from the plasma sheet thermal pressure (see N18). 

\cite{Torbert.2018} analyzed multi-probe data, including electron velocity distribution functions, and concluded that this event was consistent with simulations of laminar 2-d reconnection. N18 compared 2.5 and 3-d simulations of this reconnection event and found that near the EDR, the two simulations were nearly identical. We have determined the magnetic field dimensionality parameters of \cite{Rezeau.2018} for this event and find that they support this conclusion of \citeauthor{Torbert.2018} and N18. \cite{Rezeau.2018} defined parameters $D_1\equiv(\lambda_1-\lambda_2)/\lambda_1$, $D_2\equiv(\lambda_2-\lambda_3)/\lambda_1$, and $D_3\equiv\lambda_3/\lambda_1$, where $\lambda_1$, $\lambda_2$, and $\lambda_3$ are the eigenvalues of {the time-dependent, 3-d, symmetrix matrix $\nabla\vec{B}(\nabla\vec{B}^\mathrm{T})$ \cite{Shi.2005}}, $\lambda_1\geq\lambda_2\geq\lambda_3$, and $D_1+D_2+D_3=1$. $D_3$, which is associated three-dimensional structure, was very small as $\lambda_3$ was within the uncertainty in $\nabla\vec{B}\nabla\vec{B}^\mathrm{T}$ ($\leq10^{-5}$ nT$^2$/km$^2$). On average, throughout the EDR, we found that $D_1\approx90-97\%$ and $D_2\approx3-10\%$, implying that (1) the magnetic field gradients were much stronger in the $N$ direction that in $L$ and that (2) any gradients in the out-of-plane direction were too small to be resolved. 

During the $\sim$6-second EDR crossing (22:34:00--22:34:06 UT), MMS moved mostly laterally through the EDR in the $L$ direction while largely remaining southward of the current sheet center. Between 22:34:01--22:34:02 UT the spacecraft exited the electron current layer and crossed the separatrix into the inflow region (see magnetic field profile in figures \ref{overview}d-f and \cite{Torbert.2018}). The current sheet moved southward and the spacecraft reentered the EDR between {22:34:02--22:34:03} UT. The $B_Z$ reversal (Figure \ref{overview}f), which corresponds roughly to the crossing of the reconnection mid-plane, occurred between 22:34:02--22:34:03 UT at approximately the same time as the reversal of the electron jets (Figure \ref{overview}j). A small $\sim$2 nT Hall magnetic field was observed between 22:34:01--22:34:02.5 UT (Figure \ref{overview}e), when the spacecraft were southward of the central electron current layer and tailward of the reconnection mid-plane. An intense ($\leq$30 mV/m) northward Hall electric field (Figure \ref{overview}i) was observed by all four spacecraft throughout the EDR encounter. MMS-3, which was the only spacecraft that crossed northward of the current sheet for a significant amount of time (1--2 seconds), observed a reversal in the Hall electric field.  

\subsection{PIC simulation set-up}

We analyze the 2.5-dimensional particle-in-cell (PIC) simulation of our companion paper, N18, which used the initial conditions listed in Table 1 {(e.g., an initial plasma sheet density of $n_0=0.09$ cm$^{-3}$, an initial {background} lobe density of $n_0=0.03$ cm$^{-3}$, etc.)} to define the initial conditions of their 1-d Harris current sheet with a weak guide field ($B_G=0.03B_b$). The strength of the guide field was chosen based on the value of $B_M$ during crossings of the current sheet near the EDR, where $B_M$ was determined in the $LMN$ coordinate system based on minimum variance analysis of the electron velocity (MVA-$v_e$), as is discussed later. The simulation was created with the VPIC code \cite{Bowers.2008,Bowers.2009}. The ion-to-electron mass ratio was 400, the system size was $120d_{i0}\times40d_{i0}$ ($d_{i0}$ is the ion inertial length of the initial plasma sheet), and a total of $1.4\times10^{11}$ super-particles were used. The boundaries along the $L$ direction were periodic and the boundaries along the $N$ direction were conducting walls. Reconnection was initiated from a weak magnetic perturbation, as described in N18. As in N18, we analyze the simulation output from a single point in time 50 ion cyclotron periods after the start of the simulation ($t=50\Omega_{ci0}$) when reconnection was proceeding near the EDR in a steady-state.

For a more detailed description of the simulation set-up and choice of virtual probe path, the reader is directed to N18. In their study, N18 also determined the normalized reconnection rate $\mathcal{R}$ of their simulation by evaluating the strength of the reconnection electric field at the X-point {normalized by $v_{out}B_b$} and by determining the opening angle of the separatrix \cite{Liu.2017}. The normalized rate determined from the electric field was $\mathcal{R}$=0.17. The normalized rate determined with the method of \cite{Liu.2017} was $\mathcal{R}$=0.186.

\begin{table*}
\centering
\caption{Selected normalization parameters for PIC simulation of N18.}
\begin{tabular}{| c || c | c | c | c | c | c |}
\hline
Parameter: & $n_0$ ($n_b$) & $d_{e0}$ ($d_{eb}$) & $B_b$ & $B_G/B_b$ & $E_b=V_{Aib}B_b$ & $V_{Aib}$  \\
\hline
Value:         & 0.09 cm$^{-3}$ (0.03 cm$^{-3}$) & 17.7 km (30.7 km) & 12 nT & 0.03 & 18.12 mV/m & 1510 km/s \\
\hline
\end{tabular}
\end{table*}

\subsection{Virtual MMS data}

N18 determined an irregular cut through their simulated 2-d EDR at $t=50\Omega_{ci0}$. In their paper, they referred to this cut as ``orbit 1-s'', the $N$ coordinate of which is given by the red curve in Figure \ref{overview}p. The $L-N$ location was optimized such that the $B_L$ along the cut matched the $B_L$ observed by MMS-3, assuming that the velocity of MMS through the EDR was constant in the $L$ direction. The virtual data along this path, some of which is shown in Figure \ref{overview}m--p, reproduced many of the key features of reconnection that were observed by MMS-3 during its flight through the EDR. The small Hall magnetic field during the excursion into the inflow region, the strong and varying Hall electric field, the intense electron jet reversal and out-of-plane electron current, the strength of the normal reconnection magnetic field component, the asymmetry between the earthward and tailward electron jets, etc., are in reasonably good qualitative and quantitative agreement with the MMS-3 data.

\begin{figure*}
\noindent\includegraphics[width=22pc]{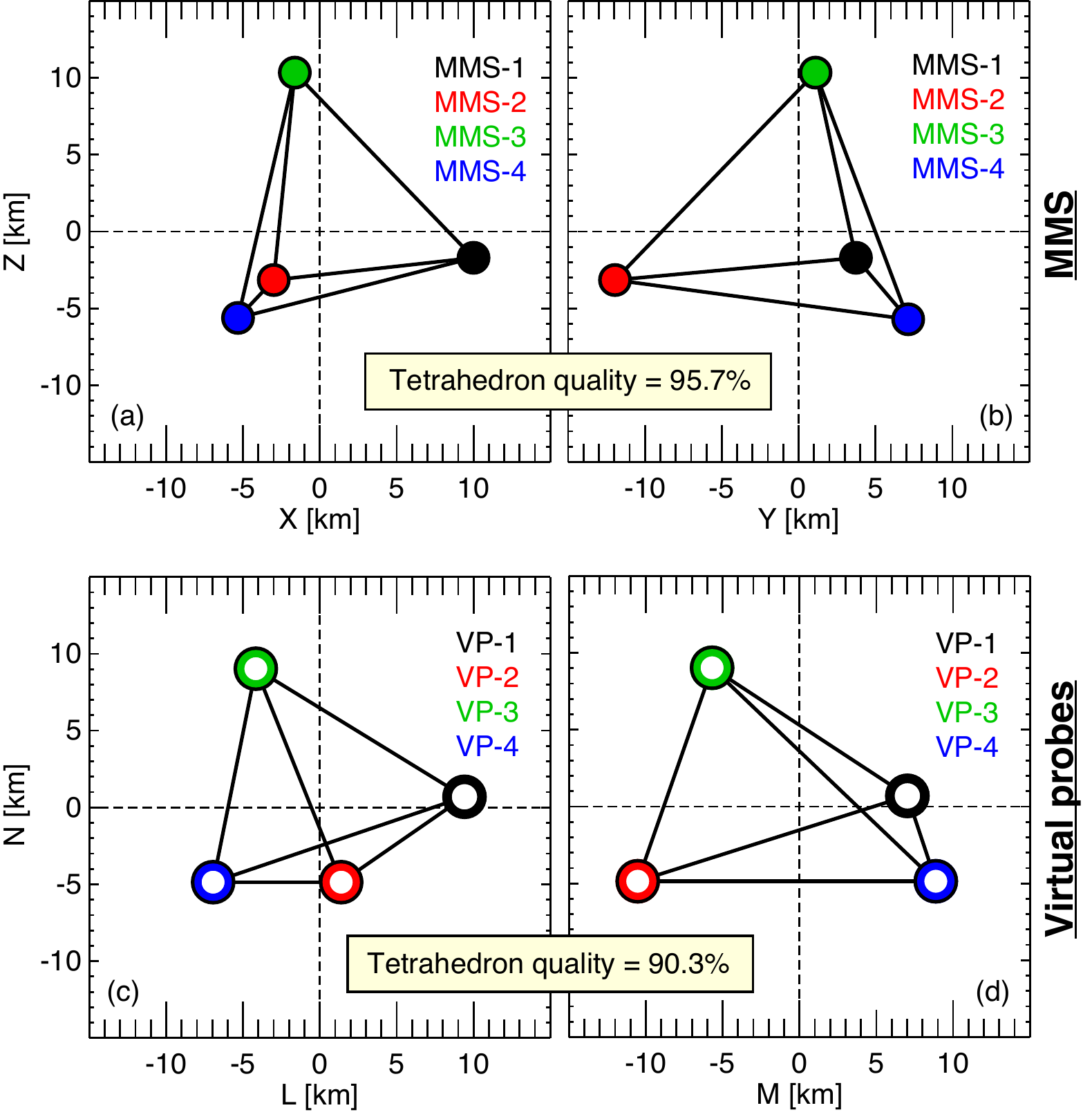}
\caption{The configuration of the MMS tetrahedron in the (a) $X-Z_{GSM}$ and (b) $Y-Z_{GSM}$ planes. The configuration of the virtual tetrahedron in the (c) $L-N$ and (d) $M-N$ planes. The tetrahedron quality factor is primarily based on the difference in volume between the actual tetrahedron and a regular tetrahedron with axes of the length of the average inter-probe distance \cite{Fuselier.2016}.}
\label{tetrahedrons}
\end{figure*}

Three other virtual probe paths were created to complete the virtual MMS tetrahedron, which were based on the path of the virtual MMS-3 orbit and the location of MMS-1, 2, and 4 relative to MMS-3 (see Figure \ref{overview}p). We then found it necessary to adjust the virtual probe positions to maintain a relatively regular tetrahedron, given that the inter-spacecraft separation of MMS (0.5--0.6$d_{e0}$) was on the order of the separation between (discrete) grid cells ($\sim$0.1 $d_{e0}$). The configurations of the MMS and virtual tetrahedrons are shown in figures \ref{tetrahedrons}a-b and in \ref{tetrahedrons}c-d, respectively. To confirm that the virtual tetrahedron was regular enough to be considered ``MMS-like'', we compared the current density vectors from the curlometer technique \cite{ISSIchap14} and 4-point-averaged plasma moments, which had a very high correlation ($R$=0.993) similar to that of MMS ($R$=0.990). The similarity of the virtual and actual tetrahedrons ensures that the errors resulting from the assumption of linear gradients during multi-point analysis should also be similar.

\section{Finding $LMN$ and $\mathcal{R}$ with MMS data}

\subsection{Defining the coordinate systems}

We have identified 14 $LMN$ coordinate systems for the 2017-07-11 EDR event using a number of different techniques, which range from the overly simple to the extremely sophisticated. The axes of the 14 coordinate systems are shown in Figure \ref{all_axes}. The coordinate axes are also tabulated in Appendix A. In general, $L$ is mostly along $X_{GSE}$, $M$ is along $Y_{GSE}$, and $N$ is along $Z_{GSE}$. The average angular separation between $L$ axes is 16$^\circ\pm11^\circ$, the average separation between $M$ axes is 19$^\circ\pm11^\circ$, and for $N$, the average separation is 13$^\circ\pm7^\circ$.

\begin{figure*}
\noindent\includegraphics[width=18pc]{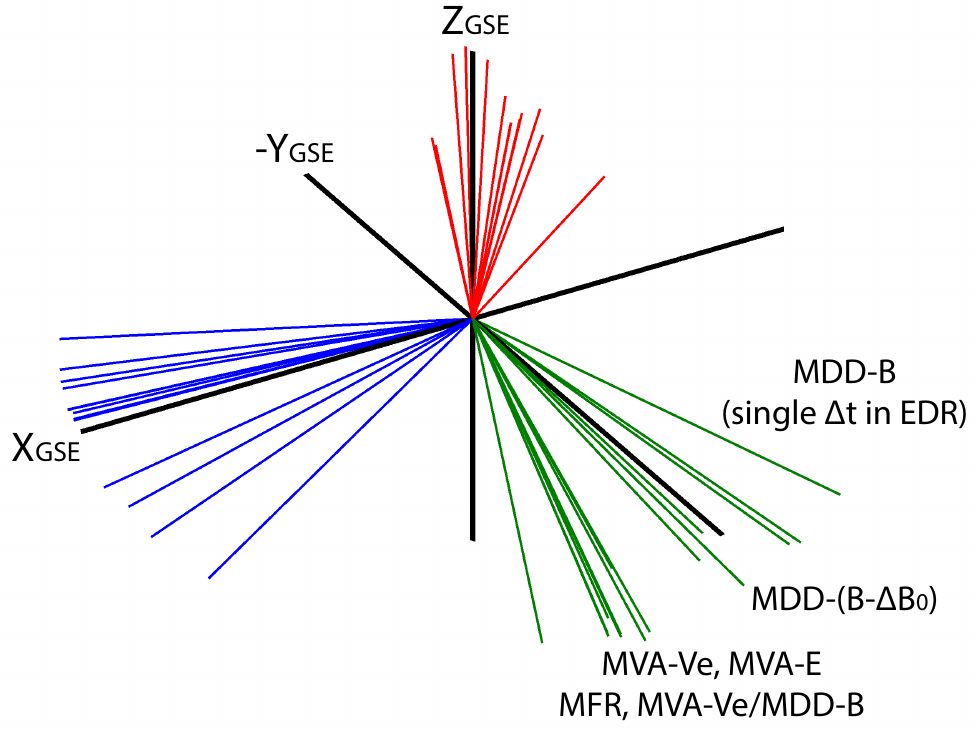}
\caption{The $LMN$ axes of all coordinate systems, where $L$ axes are colored blue, $M$ axes are green, and $N$ axes are red.}
\label{all_axes}
\end{figure*}

\subsubsection{Simple coordinates}

Our first two coordinate systems are not based on MMS data and are likely overly simple. We use solar-wind-aberrated GSM (GSW) coordinates [c.f. \textit{Fairfield}, 1980], where $L=X_{GSW}$, $M=Y_{GSW}$, and $N=Z_{GSW}$. 

We define another $LMN$ system with the empirical neutral sheet model of \cite{Fairfield.1980}, where $L=X_{GSW}$, $M$ is the normalized projection of $Y_{GSW}$ onto the current sheet surface and is perpendicular to $L$, and $N$ is the modeled current sheet normal. 

\subsubsection{Minimum variance analysis (MVA) coordinates}

The remaining coordinate systems are determined with MMS data. The following techniques define $L$, $M$, and $N$ as the vector solutions to an eigenvalue problem. To identify the ``best-quality'' coordinate system, we first select a time period over which to apply a technique. We then adjust the time period such that the eigenvalues $\lambda_L$, $\lambda_M$, and $\lambda_N$ are well separated. The coordinate axes should also be relatively stable when the time period is altered slightly.

We define two $LMN$ coordinate systems for the 2017-07-11 event using minimum variance analysis of the magnetic field (MVA-$B$) \cite{SonnerupandCahill.1967}. First, we apply MVA-$B$ to a long-duration current sheet crossing ($\sim$22:05-22:55 UT), excluding intervals where flux ropes, a moderate-strength and varying Hall magnetic field, and a weak but varying reconnection magnetic field were observed, i.e., where the current sheet is clearly not 1-d (see section 2.2 and Fig. \ref{overview}a-c). Since we exclude the interval containing the reconnection site, this technique assumes that the current sheet orientation did not change significantly in time. It also assumes that the configuration of the EDR (at the time when it observed) is identical to the average long-time-scale configuration of the current sheet. 

MVA-$B$ is also applied over a shorter time scale crossing of the current sheet near the EDR ($\sim$22:30-22:40 UT). By reducing the timespan over which MVA-$B$ is applied, any errors in the coordinate system caused by temporal or spatial variations in the current sheet orientation should be mitigated. However, since all three components of the magnetic field are expected to vary over this time interval (unlike a 1-d current sheet), the eigenvector system of the variance matrix may not represent the actual natural coordinate system of the current sheet and reconnection site.  

We define two more $LMN$ coordinate systems using MVA-$E$ \cite{Paschmann.1986,Sonnerup.1987}. {MVA-$E$ defines the $N$ and $M$ directions as the maximum and minimum viance directions of $\vec{E}$, respectively.} The first coordinate system is found by applying MVA-$E$ to the entire reconnection site interval centered on the EDR ($\sim$22:32-22:45 UT), wherein the spacecraft entered the ion-scale current sheet from the southern inflow region before exiting back into the southern inflow region. The $L$ and $M$ coordinate axes were apparently not well resolved at this time scale (i.e., $\lambda_L/\lambda_M$ was never much larger than one). This is possibly due to the large electric fields in the separatrix region observed during the current sheet crossing, errors in the current sheet velocity frame (which was assumed to be constant), time dependence effects or interactions between the reconnection site and the downstream system that lead to variations in the reconnection electric field $E_M$, etc. A better coordinate system may be defined using the joint variance technique of \cite{MozerandRetino.2007}, where MVA-$E$ is used to identify $N$ and then MVA-$B$ is used to identify $L$.

Another coordinate system is defined by applying MVA-$E$ to data from within the EDR (22:34:00.7-22:34:03.9 UT). Here, the amplitude of the bipolar $E_N$ is extremely large and reversals of $E_N$ are observed with each partial current sheet crossing (Fig. \ref{overview}i). Also in the EDR, there is a moderate $E_L$ that reverses its polarity near the $B_N$ reversal (Fig. \ref{overview}g,f). Finally, since only a short timespan around the central EDR is considered, it is likely more reasonable to assume that the reconnection electric field $E_M$ should vary minimally here. Unlike the previous MVA-$E$ coordinate system, $\lambda_L/\lambda_M$ was large (=32.7) and $\lambda_N/\lambda_L$ was poor (=2.9). For other time intervals within the EDR, both $\lambda_L/\lambda_M$ and $\lambda_N/\lambda_L$ were moderate. Since the focus of this paper is the reconnection rate $E_M$, however, we chose to maximize the quality of $M$.

Two more $LMN$ coordinate systems are defined by applying MVA to the ion and electron bulk velocities, MVA-$v_i$ and MVA-$v_e$, respectively. For both of these coordinate systems, it is assumed that $L$ is the jet reversal and thus the maximum variance direction, $N$ is the inflow direction and thus the minimum variance direction, and $M$ is the intermediate variance direction. The MVA-$v_i$-based coordinate system was determined over the ion jet reversal period and was reasonably stable. However, the eigenvalue resolution was poor ($\lambda_L/\lambda_M\approx4$ and $\lambda_M/\lambda_N\approx6$) and the coordinate system did not organize the data near the EDR. This may be due to the quality of the ion moments data (see discussion in \textit{Torbert et al} [2017]) or possibly due to asymmetries in the jets and/or non-uniform cross-tail current structure. 

Another coordinate system was determined by applying MVA-$v_e$ to MMS-3 data from the electron jet reversal period in the central EDR (22:34:02-22:34:04 UT). Again, the eigenvalue resolution was poor ($\lambda_L/\lambda_M=4.4$ and $\lambda_M/\lambda_N=14$), though the resulting coordinate system seemed to do a very good job at organizing the data in and around the EDR (see discussion and Figure \ref{datainlmn}a-c in the next section). The poor eigenvalue resolution may be due to the considerable amount of noise in $v_e$, which might affect MVA by adding unphysical variance. Our possibly naive attempts to filter out the noise (boxcar averaging, smoothing, etc.) moderately improved the eigenvalue separation, but appeared to reduce the quality of the coordinate system (one example being that the reconnection electric field became time-varying, at times unrealistically large or strongly negative). We also note that the eigenvalue separation and coordinate system quality was reduced when MVA-$v_e$ was applied to the three spacecraft that were further from the current sheet center. This point will be examined in section 4 with virtual data from our simulation. Lastly, we note that, since the density across the EDR is almost constant (excepting noise) and $v_i<<v_e$, MVA-$v_e$ is essentially identical to MVA-$J$.

\subsubsection{Minimization of Faraday residue (MFR) coordinates}

Another $LMN$ coordinate system is defined with the minimization of Faraday residue (MFR) technique \cite{KhrabrovandSonnerup.1998}. In MFR, the coordinate axes are found from time series data of $\vec{E}$ and $\vec{B}$ from a single spacecraft and are coupled to the velocity of the boundary layer along its normal. We found that the coordinate axes were unstable when changes were made to the time interval over which MFR was applied, possibly due to the irregular and time-dependent EDR motion (see predicted path of MMS in Figure \ref{overview}p) and/or the complex structure of the current sheet at the EDR \cite{Sonnerup.2006}. However, the eigenvalue separation reached a clear maximum for the period between 22:34:02 and 22:34:03.5 UT ($\lambda_L/\lambda_M=6.2$ and $\lambda_M/\lambda_N=50.1$). The MFR normal velocity of the current sheet was $u_N=86.6$ km/s, which is reasonably close to the normal velocity of $\sim70$ km/s that was obtained by \cite{Torbert.2018} via timing analysis of the $B_N$ reversal. We have also applied the method of \cite{SonnerupandHasegawa.2005}, which is essentially a generalization of MFR for a 2-d boundary layer. We do not find any period near the EDR over which this method returns sensible and stable results, which may be due to the irregular motion of the EDR.

\subsubsection{Maximum directional derivative of $B$ (MDD-$B$) coordinates}

Lastly, we define a group of $LMN$ coordinate systems based on the maximum directional derivative of $\vec{B}$ (MDD-$B$) technique of \cite{Shi.2005}. This technique can be used to find a coordinate system for every point in time where 4-point measurements of $\vec{B}$ are made. The logic is as follows: $N$ is the direction along which the gradient of $\vec{B}$ is maximized, $M$ is the invariant direction of $\vec{B}$, and $L$ is the intermediate gradient direction. A time-varying $LMN$ coordinate system is then defined by the eigenvectors of the symmetric{, time-dependent 3x3} matrix $\nabla\vec{B}(t)(\nabla\vec{B}(t))^\mathrm{T}$. An average coordinate system can be defined by finding the eigenvectors of $\left<\nabla\vec{B}(t)(\nabla\vec{B}(t))^\mathrm{T}\right>$ after averaging the matrix over some period of time where the time-varying axes are stable.  The measured electron velocities (Figure \ref{overview}j-l) and current densities are very similar for all four spacecraft, which implies that the magnetic field likely varies linearly within the spacecraft tetrahedron.

First, we find $L$, $M$, and $N$ simultaneously by applying MDD-$B$ to data from the period between 22:34:02-22:34:03.1 UT. The ratio of the $N$ and $M$ eigenvalues is large ($\lambda_N/\lambda_M=752$), but the data are most likely very poorly organized by the resulting coordinate system, as will be discussed in section 3.2. 

Next, we define a coordinate by applying MDD-$B$ to two different periods, one for which $M$ is stable and well resolved (22:34:01.6-22:34:03.1 UT) and then another for which $N$ is stable and well resolved (22:34:02.3-22:34:04 UT). The $N$ axis is then adjusted to be perpendicular to $M$. The $L$ axis is defined by their cross product. 

We define yet another coordinate system where the inter-calibration of the 4-point measurements of $\vec{B}$ is adjusted before applying MDD-$B$. We first calculate the average differences between the 4-probe magnetic field vectors $\left<\Delta\vec{B}\right>_0$ over a quiet 2-minute period (22:49-22:51 UT). In this quiet interval, the curlometer current was much larger than the current detected by the plasma instruments, implying that the gradients in $\vec{B}$ should be {largely unphysical}. We find that the average values of $\left<\Delta\vec{B}\right>_0$ for each vector component for each spacecraft are smaller than $\sim$0.05 nT, which is well within the reported error for the magnetometer data (see section 2.1). We also find that subtracting $\left<\Delta\vec{B}\right>_0$ from $\vec{B}$ reduces the value of the linear approximation of $\nabla\cdot\vec{B}$ in the interval around the EDR. ($\nabla\cdot\vec{B}$ is commonly associated with the error in the linear gradient technique, though, in this case, it is likely associated with small errors in the inter-calibration of the magnetometers). There is still an apparent residual spin-tone of $\sim\pm0.02$ nT in the spin-plane components of $\vec{B}$, but we get poor results from fitting and extrapolating this spin-tone beyond the quiet-time interval. The adjusted MDD-$B$ coordinate system (referred to as MDD-($B-\Delta B_0$)) is defined in an almost identical manner to the last coordinate system, though the MDD-$B$ matrix is found by $\nabla(\vec{B}-\left<\Delta\vec{B}\right>_0)(\nabla(\vec{B}-\left<\Delta\vec{B}\right>_0))^\mathrm{T}$. This technique is similar to the ``perturbed MDD-$B$'' technique of \cite{Denton.2010,Denton.2012}. While this method is expected to account for constant errors in $\vec{B}$, it does not account for the time-dependent spin tone in the spin-plane components of $\vec{B}$.

\subsubsection{Hybrid coordinate systems}

The final two coordinate systems are based on hybrid techniques, where $N$ is determined from MDD-$B$ and the other directions are determined separately. Similar coordinate systems were determined with MMS data by \cite{Denton.2018}. One coordinate system is from \cite{Torbert.2018}, which applied MDD-$B$ to data at the $B_Z$ reversal to determine $N$, used the maximum component of the time-averaged current to determine $M$, and then found $L$ to complete the right-handed coordinate system. 

For our final coordinate system, $L$ was defined with MVA-$v_e$, $M$ was defined as the cross product of $L$ and the normal from MDD-$B$, and $N$ completed the right-handed system.

Other multi-probe techniques for finding the normal direction, e.g., constant velocity or timing analysis (CVA or TA) \cite{Haaland.2004} and local normal analysis \cite{Rezeau.2018}, may be used to find additional hybrid $LMN$ coordinate systems. These techniques were also applied to the 2017-07-11 event; however, issues related to the data quality and/or crossing geometry prevented these techniques from producing reasonable normal directions.

\subsubsection{Summary of 14 $LMN$ coordinate systems}

In summary, 14 coordinate systems are found by:
\begin{enumerate}
\item Using GSW coordinates,
\item Using modified GSW coordinates fixed to the \cite{Fairfield.1980} neutral sheet model,
\item Applying MVA-$B$ to data from a long-duration current sheet crossing, excluding the reconnection region,
\item Applying MVA-$B$ over the reconnection region,
\item Applying MVA-$E$ over a long time interval with a current sheet crossing,
\item Applying MVA-$E$ over a short time interval surrounding the EDR,
\item Applying MVA-$v_i$ over the ion jet reversal,
\item Applying MVA-$v_e$ over the electron jet reversal,
\item Applying MFR within the EDR, over a period where the normal velocity appeared to be steady,
\item Applying MDD-$B$ over one time period,
\item Applying MDD-$B$ over two time periods (one to find $M$ then another to find $N$ and thus $L$),
\item Applying MDD-$(B-\Delta B_0)$ over the same two time periods after subtracting {the small average inter-probe differences $\left<\Delta\vec{B}\right>_0$} from each of the four measurements of $\vec{B}$,
\item Defining $M$ as the maximum direction of the current density in the EDR, applying MDD-$B$ near the X-point to find $N$ perpendicular to $M$, and defining $L$ perpendicular to $M$ and $N$ (see \cite{Torbert.2018}), and finally
\item Defining $L$ with MVA-$v_e$, defining $M$ as the cross product of $L$ and the MDD-$B$ normal, and finding $N$ perpendicular to $L$ and $M$.
\end{enumerate}

\subsection{The EDR structure in different coordinate systems}

\begin{figure*}
\noindent\includegraphics[width=39pc]{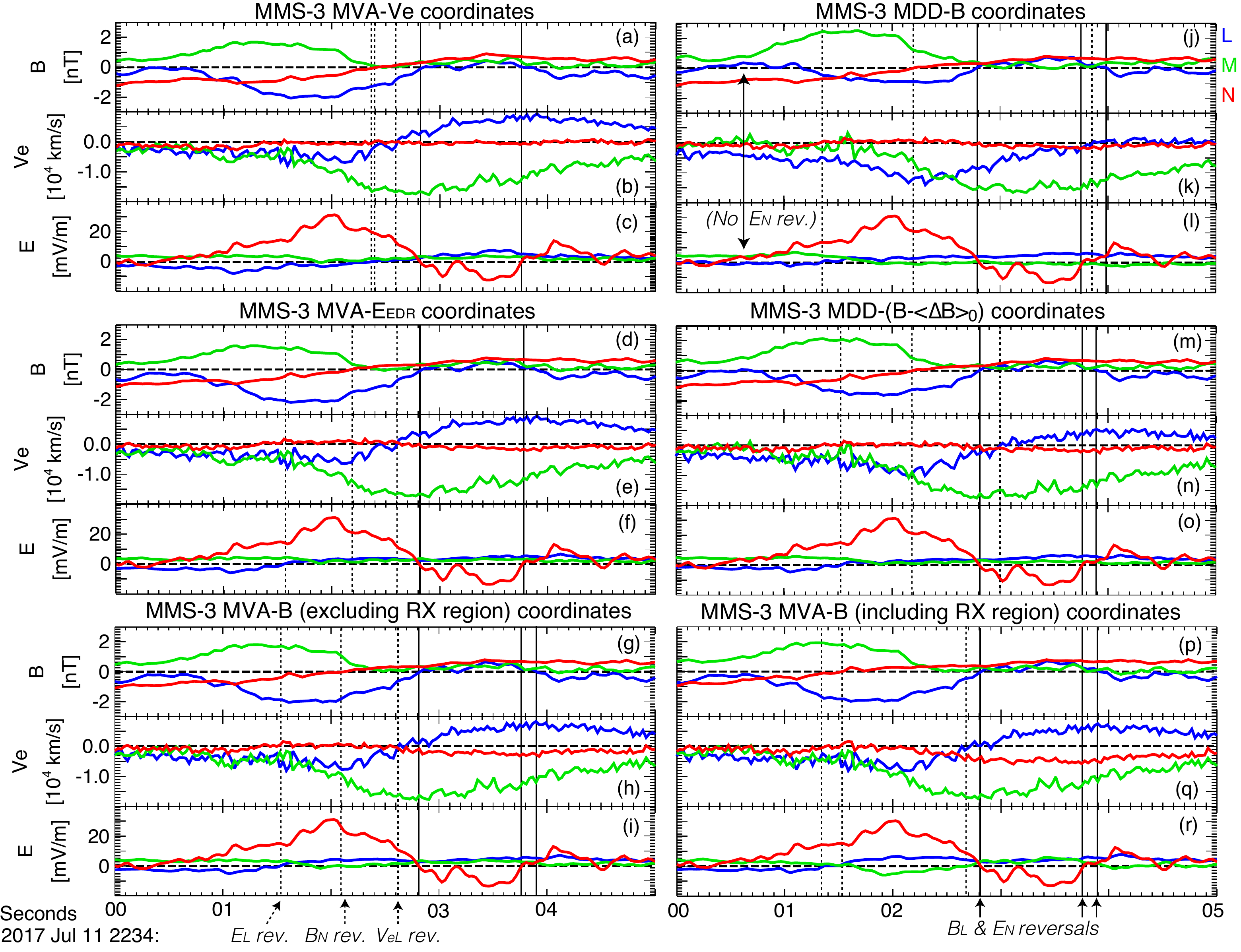}
\caption{MMS-3 data, which are shown in the $LMN$ coordinates that were determined by (a)-(c) applying MVA-$v_e$ to data from 22:34:02-22:34:04 UT, (d)-(f) applying MVA-$E$ to data from 22:34:00.7-22:34:03.9 UT, (g)-(i) applying MVA-$B$ to data from the interval $\sim$22:05-22:55 UT (excluding the reconnection region around $\sim$22:30-22:40 UT where the current sheet is clearly not 1-d),  (j)-(l) applying MDD-$B$ to data from 22:34:02-22:34:04 UT, (m)-(o) applying MDD-$(B-\Delta B_0)$ to data from different time periods in which $L$, $M$, and $N$ are individually stable and subtracting the average magnetic field gradient determined over a quiet interval before finding the MDD-$B$ matrix, and (p)-(r) applying MVA-$B$ to data from the interval $\sim$22:30-22:40 UT. The vertical dashed lines mark the reversals of $E_L$, $B_N$, and $v_{eL}$, which should be aligned in time according to our model. The vertical solid lines mark the reversals of $B_L$ and $E_N$, which were observed after the mid-plane crossing.}
\label{datainlmn}
\end{figure*}

Figure \ref{datainlmn} shows $\vec{B}$, $\vec{v}_e$, and $\vec{E}$ data from MMS-3 during the EDR observation. The data are shown in six different coordinate systems, which are listed in the figure caption. We compare the data in these coordinate systems to what would be expected for a simple picture of 2-d, steady-state, laminar, and symmetric reconnection, which seems to be a reasonable approximation for this event, as is discussed in section 2.2 and N18. The vertical dashed lines mark the reversals of $B_N$, $v_{eL}$, and $E_L$, which are expected signatures of a crossing of the reconnection mid-plane. These signatures are expected to be simultaneous for the simple reconnection picture. The solid vertical lines mark the major reversals of $B_L$ and $E_N$, which are signatures of a neutral sheet crossing. Away from the neutral sheet, $E_N$ and $B_L$ should be oppositely signed for anti-parallel reconnection; however, as is shown by our virtual data in Figure \ref{overview}m and n, like signs of $E_N$ and $B_L$ (and $B_N$) may occur with even a very small guide field. Specifically, for a small positive guide field, $E_N$ may be negative tailward and immediately southward of the neutral sheet (i.e., $E_N$, $B_N$, and $B_L$ are all small and negative) whereas $E_N$ may be positive earthward and immediately northward of the neutral sheet (i.e., $E_N$, $B_N$, and $B_L$ are all small and positive). This is also shown in Figure 6 of our companion paper, N18. Also in the simple reconnection picture, the reconnection electric field $E_M$ {should be uniform and positive around the diffusion region, given $E_M$ is responsible for the steady circulation and change of connectivity of flux tubes in the EDR. Lastly, the normal electron bulk velocity $v_{eN}$ should be very small or zero at the neutral sheet where ($B_L$=0), given that the momenta of two symmetric inflow regions balance one another at the current sheet center.}

As is evident in Figure \ref{datainlmn}a-f, the data in the MVA-$v_e$ and MVA-$E$ coordinate systems look very similar. The most pronounced crossing of the neutral sheet occurs between 22:34:02.8--22:34:03.8 UT and has nearly simultaneous reversals in $E_N$ and $B_L$. The guide field, as defined by the strength of $B_M$ at the X-point, is $\sim$0.3-0.4 nT for both coordinate systems. $E_M$ is relatively constant, small, and positive in both coordinate systems. This condition is used to define the MVA-$E$ coordinate system, but it is not considered during MVA-$v_{e}$. In the MVA-$v_e$ coordinate system, the reversals of $B_N$, $v_{eL}$, and $E_L$ occur within a quarter of a second. For the MVA-$E$ coordinate system, wherein the quality of $M$ was preferred over $N$ and $L$, these reversals are observed within roughly one second of each other. Lastly, we note that $v_{eN}$ is either very small or zero at neutral sheet in both the MVA-$v_e$ and MVA-$E$ coordinate systems. This is a condition used to define the MVA-$v_e$ coordinate system, but it is not considered during MVA-$E$. (Though this is not pictured here, the MFR and hybrid MVA-$v_e$/MDD-$B$ coordinate systems are quite similar to the MVA-$E$ and MVA-$v_e$ coordinate systems).

The MDD-$B$ coordinate system, in which the data in Figure \ref{datainlmn}g-i are shown, is quite different from the MVA-$v_e$ and MVA-$E$ systems, as is also shown in Figure \ref{all_axes}. The electron jet is highly asymmetric. The earthward jet is barely visible above the noise. The reversals in $B_N$, $v_{eL}$, and $E_L$ are separated from one another by 2.5 seconds (compared to $<0.25$ seconds for MVA-$v_e$ and $<$1 second for MVA-$E$). Nearly simultaneous reversals in $B_L$ and $E_N$ are observed after the $B_N$ reversal, but there is a nearly 1-second reversal of $B_L$ around 22:34:00.5 UT that is not associated with any significant reversal in $E_N$. The signs of $B_L$, $B_N$, and $E_N$ here do not match our picture of weak guide-field $B_M>0$ reconnection. The guide field strength estimated in these coordinates is $B_G>1$ nT, which is significantly larger than previously expected but still small compared to the $B_b=$12 nT background field. Lastly, the reconnection electric field is not constant and often negative. The MDD-$(B-\Delta B_0)$ coordinate system (Figure \ref{datainlmn}j-l) compares much more favorably with the simple reconnection picture and with the data in the MVA-$v_e$ and MVA-$E$ coordinate systems. For instance, {the subtraction of $\Delta B_0$} (1) reduces the time between the $v_{eL}$, $B_N$, and $E_L$ reversals by a factor of 2.5, (2) leads to $E_M$ remaining small, relatively constant, and positive, (3) enhances the asymmetry in $E_L$ in a manner that matches our virtual data, (4) reduces the asymmetry in $v_{eL}$ in a manner that matches our virtual data, (4) reduces the guide field strength to $B_G\approx0.5$ nT, which is comparable to its determined value with MVA-$v_e$ coordinates, etc. The $L$, $M$, and $N$ axes of the MDD-($B-\Delta B_0$) systems are separated by $21^\circ$, $21^\circ$, and $2^\circ$ from the corresponding axes of the original MDD-$B$ system. 

Lastly, we consider MVA-$B$. Both of our MVA-$B$ coordinate systems are determined over a much larger timespan than any of the others shown in Figure \ref{datainlmn}, since a full current sheet crossing is required for this technique. Proper identification of $L$, $M$, and $N$ by MVA-$B$ requires the current sheet to be roughly 1-d. Since the timespans for these coordinate systems are much longer duration than the EDR encounter, the resulting $LMN$ coordinates are only relevant to the EDR interval if the current sheet orientation does not change in time. Neither coordinate system matches each of the criteria laid out previously for simple 2-d, laminar, symmetric, and steady-state reconnection, though the coordinate system in Figure \ref{datainlmn}g-i (where MVA-$B$ is applied to a longer duration current sheet crossing and the reconnection region interval is excluded) arguably comes closer to doing so than the coordinate system in Figure \ref{datainlmn}p-r. This claim is based on the larger $v_{eN}$ in \ref{datainlmn}q, the longer interval of more strongly negative $E_M$ in \ref{datainlmn}r, the greater separation between the reversals of $B_N$, $E_L$, and $v_{eL}$ in \ref{datainlmn}p-r, etc. Similar to the MDD-$B$ coordinates, we take this as an indication that a more educated application of a coordinate system technique tends to make the MMS data more closely resemble both the virtual data and the simple reconnection picture.

\subsection{Calculating $\mathcal{R}$ with MMS data}

Having identified 14 different $LMN$ coordinate systems, we now find $E_M$ and normalize by $E_b=18.12$ mV/m (see discussion in section 2 and N18) to obtain $\mathcal{R}$. The reconnection electric field is determined by averaging $E_M$ over the period from 22:34:03--22:34:04 UT, which is the period nearest the $B_N$ reversal where the total electric field is smallest (see section 2.2). Given that {the largest value of} $E_N$ is observed during the $B_N$ reversal, the interval around the {$B_N$ reversal} is not the ideal time to find $\mathcal{R}$, as any finite projection of the inaccurately measured $M^\ast$ onto $N$ will produce very large errors in $E_{M\ast}$. Indeed, a deflection of $E_M$ relative to its average value is observed near 22:34:02 UT, where $E_N$ is sharply peaked. For comparison, the $B_N$ reversal occurs at 22:34:02.1--22:34:02.4 UT in most coordinate systems. The time we have used to find $E_M$, 22:34:03--22:34:04 UT, also has very weak magnetic fields, meaning that the $\vec{V}\times\vec{B}$ offset in the spacecraft-frame electric field from the relative motion of the X-line is minimized. For our simple picture of 2-d, laminar, and steady-state reconnection, the reconnection electric field should be more-or-less constant in time and space, at least in the highly local region surrounding the EDR, while $E_L$ and $E_N$ are not constant and vary considerably. Therefore, we assume that a quality measurement of $E_M$ should be one which deviates minimally from its average value. We also compare our measurements of $\mathcal{R}$ with those of \cite{Torbert.2018} and N18. \cite{Torbert.2018} determined that the aspect ratio of the diffusion region was between 0.1 and 0.2. N18 determined that the opening angle of the separatrix was $\sim$12.5$^\circ$, which corresponds to a normalized reconnection rate of $\mathcal{R}\approx0.18$. These normalized reconnection rate measurements did not depend on either the magnitude of $E_M$ or $E_b$, though they are not without their own sources of error. Lastly, N18 also noted that the reconnection rate of their simulation was also $\sim0.17-0.186${, which they determined by analyzing the separatrix opening angle and normalized reconnection electric field strength with the virtual probe data (see sections 4.1, section 4.2, and Figure 9 of N18 for more details on the simulated reconnection rates).}

\begin{figure*}
\noindent\includegraphics[width=39pc]{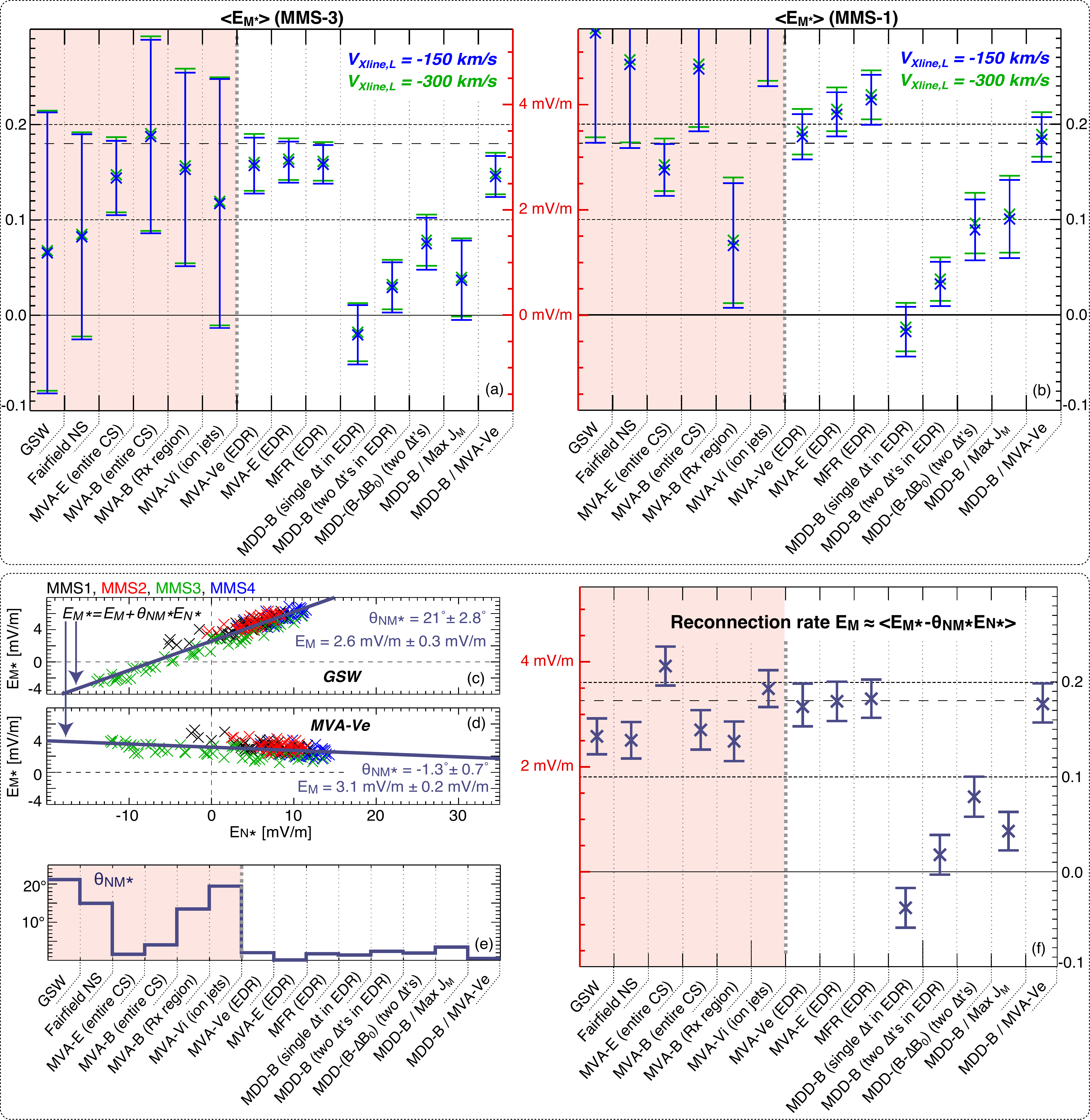}
\caption{{(a)-(b)} The reconnection rate $E_M$ in the X-line frame determined from MMS-3 (left) and MMS-1 (right) for each of the 14 $LMN$ coordinate systems. The ``X''s mark the averaged $E_M$ determined over the period 22:34:03--22:34:04 UT. The error bars mark $\pm\sigma_{E_M}$, the standard deviation of the reconnection rate over this period. {The blue ``X''s mark the reconnection rates determined in the X-line frame of \cite{Torbert.2018} and the green ``X''s are determined in a frame moving twice as fast}. The reconnection rate determined from the MMS data at the near-EDR separatrix by N18 is marked by the long-dashed horizontal line ($\mathcal{R}$=0.18). The range for the reconnection rate determined by \cite{Torbert.2018} is between the two horizontal dotted lines ($0.1\leq\mathcal{R}\leq0.2$). The data in the red shaded region are determined using coordinate systems that are not solely based on MMS data from within the EDR.}
\label{rrates}
\end{figure*}

Our 14 estimates of $\mathcal{R}$ are shown in Figure \ref{rrates}. {Figure \ref{rrates}a} shows the values of $\mathcal{R}$ measured by MMS-3, which was closer to the current sheet center than MMS-1. The reconnection rate measured by MMS-1 is shown in {Figure \ref{rrates}b}. There is a fair amount of scatter in $\mathcal{R}$ from one coordinate system to another. For some coordinate systems, $\mathcal{R}$ measured by MMS-3 also differs significantly from MMS-1. While the reconnection electric field should be more-or-less constant, this is not the case for the normal electric field, which {is, on average, four times stronger for MMS-1 than MMS-3 during 22:34:03--22:34:04 UT} (see Figure \ref{overview}i). If the $M$ axis is measured incorrectly as $M^\ast$, where $M^\ast$ has a finite projection onto $N$, then we would expect significant differences in the measured reconnection rate between MMS-1 and MMS-3. Some of the differences between the values of $E_M$ obtained by MMS-1 and MMS-3 data may also be explained by differences in the calibration of the two probes. However, given that the two values of $E_M$ are nearly the same for some of the more reliable coordinate systems (e.g., MVA-$v_e$, MVA-$E$, {and MDD-$B$/MVA-$v_e$}), the inter-calibration of the probes is not likely to be the cause for the large differences between the two values of $E_M$ in, for example, MVA-$v_i$ or GSW coordinates. We also note that the $-\vec{V}\times\vec{B}$ electric field due to the relative motion of the X-line and the spacecraft was a negligible source of error for $\mathcal{R}$, as is demonstrated by the very small differences between the blue ($E_M$ in the X-line frame {of \cite{Torbert.2018}}, assuming a tailward X-line velocity of 150 km/s) and green (tailward X-line velocity of 300 km/s{, which corresponds to a 100\% error in the X-line frame of \cite{Torbert.2018}}) crosses in Figure \ref{rrates}{a-b. Given an average value of $\left<B_N\right>\approx0.5-0.7$ nT in the period when $E_M$ is calculated, even such a large 100\% uncertainty in the X-line velocity (150-300 km/s) only corresponds to a $\sim$2-4\% error in the reconnection rate for $\left<E_M\right>\approx3$ mV/m.}

When only MMS-3 data are considered, there is {an apparent} consensus between the reconnection rates in the coordinate systems determined by (1) applying MVA-$v_e$ to the electron jet reversal, (2) applying MVA-$E$ to the EDR, (3) applying MFR to the EDR, and (4) a hybrid of MDD-$B$ and MVA-$v_e$. {Using MMS-3 data, the reconnection rates in these four coordinate systems are, respectively, 0.16, 0.16, 0.15, and 0.16. However, the reconnection rates determined using MMS-1 data from the same interval in the same four coordinate systems are, respectively, 13\%, 26\%, 38\%, and 23\% larger than for MMS-3. It is possible to use the correlation between $E_{M\ast}$ and $E_{N\ast}$ to obtain error bars for $\mathcal{R}$ if they are caused by errors of the form $E_{M\ast}=\sin(\theta_{NM\ast})E_M+\cos(\theta_{NM\ast})E_N$. In this case, the errors can be reduced to $E_{M*}\approx E_M+\theta_{NM\ast}E_{N\ast}$ using the approximations that $\theta_{NM\ast}<<1$ and $E_M^2<<E_N^2$. Since both $E_{M\ast}$ and $E_{N\ast}$ are known quantities (the values of $E_M$ and $E_N$ in any imprecise coordinate system), the values of $E_M$ and $\theta_{NM\ast}$ can be approximated by fitting a linear function to $E_{N\ast}$ versus $E_{M\ast}$. 

Fit lines of the type $E_M\ast=E_M+\theta_{NM\ast}E_N\ast$ are shown in Figure \ref{rrates}c-d for the GSW and MVA-$v_e$ coordinate systems, respectively. Unsurprisingly, we find evidence that the $M$ direction defined by the $Y_{GSW}$ axis is non-orthogonal with $N$, as shown by the strong correlation between $E_{M\ast}$ and $E_{N\ast}$. The form of the fit line can be rearranged as $\Delta E_M\equiv \left|E_{M\ast}-E_M\right|/E_M=\left|\theta_{NM\ast}E_{N\ast}\right|/E_M$ such the relative error in $E_M$ can be expressed by $\theta_{NM\ast}$ as a percentage of $E_N$. Given the maximum value of $E_{N\ast}$, which is observed to be roughly 10 times as large as $E_M$ at the $B_N$ reversal (see Figure 4), even the small error angle of $\theta_{NM\ast}=1.3^\circ$ shown in Figure \ref{rrates}d corresponds to an error of $\sim$20\%. For GSW, which had $\theta_{NM\ast}=21^\circ$, a $\sim$350\% error would be expected if $E_M$ was measured during the period of largest $E_N$. The values of $\theta_{NM\ast}$ for all 14 coordinate systems are shown in Figure \ref{rrates}e. Unsurprisingly, the value of $\theta_{NM\ast}$ is small for MVA-$E$, which defines $M^\ast$ as the direction of minimum electric field variance.

The values of $E_M$ determined by this linear regression correction are shown in Figure \ref{rrates}f. The four coordinate systems mentioned previously (MVA-$v_e$ applied to the electron jet reversal interval, MVA-$E$ applied to the EDR current sheet crossing interval, MFR applied to same interval, and a hybrid of MDD-$B$ and MVA-$v_e$) have nearly identical values of $E_M/E_b$ equal to 0.176, 0.184, 0.186, and 0.176, respectively, which have $2\sigma$ errors of $\approx$10\%-15\%. Note that this correction does not account for all sources of error related to the selection of coordinates, as (1) errors due to the non-orthogonality of $L$ and $M^\ast$ can also significantly affect the reconnection rate in a similar manner to our previous approximation with $\theta_{NM\ast}E_{N\ast}$ and (2) most of the approximations described above do not hold when multiple rotations about different axes are needed to account for finite projections of $M^\ast$ onto both $N$ and $L$. Given that this $\theta_{NM\ast}$ correction does not noticeably affect the reconnection rates determined with MDD-$B$, yet the normal direction determined from MDD-$B$ is separated by nearly 9$^\circ$ from the MVA-$v_e$ normal and $E_{M\ast}$ and $E_{L\ast}$ are not very well correlated, it is likely that correcting the MDD-$B$ reconnection rate would require more than our simple approximate linear determination of a single error angle.
}

\section{Finding $LMN$ and $\mathcal{R}$ with virtual data}

Now we estimate the errors in the measured coordinate axes and $\mathcal{R}$ from the virtual MMS data described in section 2.4 and N18. We focus on MVA-$v_{e}$ and MDD-$B$. Ultimately, our goal is to make the virtual data as MMS-like as possible so these errors are realistic. However, since we do not know, for instance, how 3-d and time-dependent effects are manifested in the MMS-data (or the degree to which they are present), the errors we estimate here will inevitably be conservative. However, as was discussed in sections 2 and 3.2, if 3-d and time-dependent effects are indeed manifested in the MMS-data, they do not seem to cause any major differences between the actual and virtual MMS data.

First, we consider the errors in the MVA-$v_e$ coordinate system. Assuming that the direction of $v_e$ is correct on average, the only source of measurement error should be the noise. We have estimated that the noise in $v_e$ is around 2500 km/s, which is 14\% of the largest value of $v_{eM}\approx15000$ km/s observed by MMS-3 in the EDR. Also, in the virtual MMS-3 data (Figure \ref{overview}o), $v_{eN}$ varies over the electron jet reversal period, though MVA-$v_e$ defines the $N$ direction as the direction of minimum variance. Lastly, we reiterate that the separation between the $L$ and $M$ eigenvalues was larger for MMS-3 than it was for the other three spacecraft. In total, we conclude that the most likely candidates for sources of error in MVA-$v_e$ are noise and incorrect assumptions about the configuration of the electron velocity in $LMN$, which may be worsened with distance to the current sheet center.

\begin{figure}
\noindent\includegraphics[width=14pc]{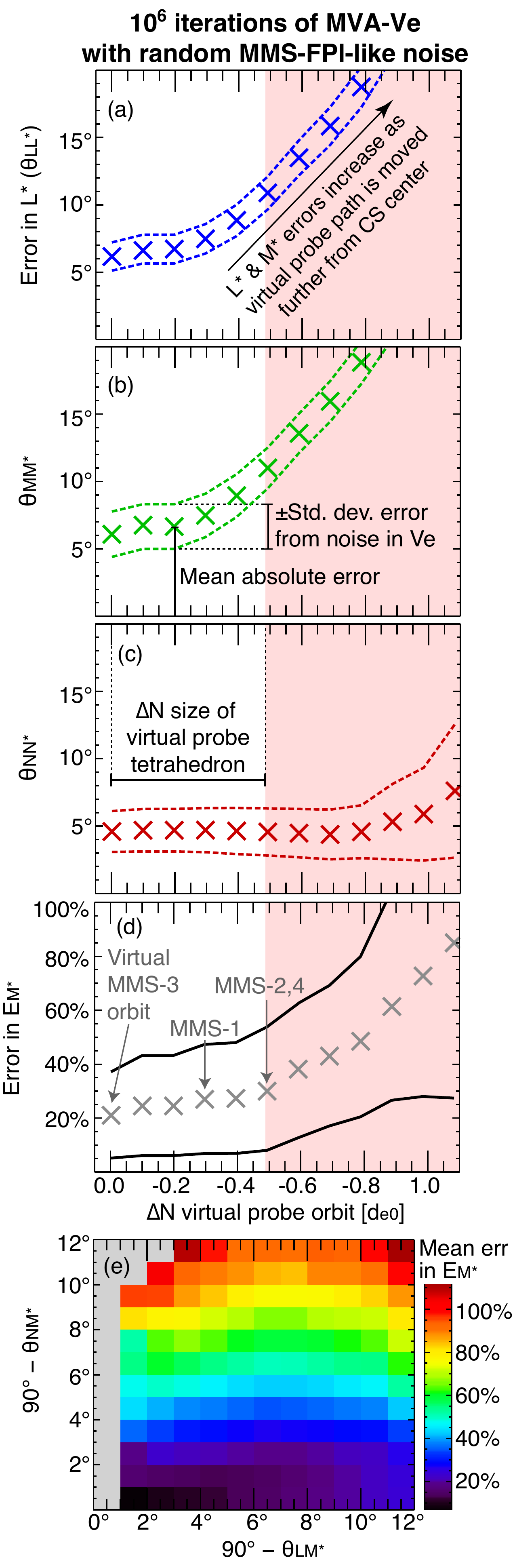}
\caption{Errors in the coordinate axes and reconnection rate determined by applying MVA to the virtual electron bulk velocity $v_e$ after random ``MMS-like'' noise was added. (a)-(d) Show these errors as a function of the distance of our MMS-like virtual probe path from its initial location near the current sheet. The ``X'' marks indicate the average errors from $10^6$ iterations. The dashed lines are the average values plus or minus a standard deviation. The vertical axes are (a): the error in the measured $L^\ast$ axis, $\theta_{LL^\ast}\equiv\cos^{-1}\left(\hat{L}\cdot\hat{L}^\ast\right)$, (b): $\theta_{MM\ast}$, (c): $\theta_{NN\ast}$, and (d): the percent error in $E_{M\ast}$. The red-shaded box indicates the region where the displacement in the virtual probe orbit is larger than the tetrahedron size. Panel (e) shows the degree of non-orthogonality between $L$ and $M^\ast$ (|$90^\circ-\theta_{LM\ast}$|) on its horizontal axis, |$90^\circ-\theta_{NM\ast}$| on the vertical, and the mean error in the reconnection rate per $1^\circ\times1^\circ$ bin in color.}
\label{mvave_errs}
\end{figure}

Figure \ref{mvave_errs} shows the error in the coordinate axes, $L^\ast$, $M^\ast$, and $N^\ast$, as well as $\mathcal{R}^\ast$, which were determined by applying MVA to the noisy virtual $v_e$. As in section 1, we use the asterisk to mark quantities that are known to be incorrect. For example, $L^\ast$ may be defined by the maximum variance direction of $v_e$; however, it is known to be different from the true $L$ axis of the simulation. The total angular error in $L^\ast$ is referred to as $\theta_{LL^\ast}\equiv\cos^{-1}(\hat{L}\cdot\hat{L}^\ast)$, being the angle between the measured $L^\ast$ axis and the true $L$ axis. To find the error terms that are shown in Figure \ref{mvave_errs}, we have done the following: (1) we averaged $v_e$ along the virtual probe path in order to obtain a realistic number of data points ($\sim$70) within the electron jet reversal interval, such that the resolution of the virtual data is comparable to the resolution of FPI. (2) We introduced noise to $v_e$ using a random number generator. The most probable value for the random noise is 0 km/s and the standard deviation of the noise was chosen to be $\pm14\%$ of the largest value of $v_e$ along the path of the virtual MMS-3 (see red curve in Figure \ref{overview}p). (3) We adjusted the interval to maximize the eigenvalue separation. (4) We applied minimum variance analysis to the noisy $v_e$ data, reiterating the process $10^6$ times to ensure statistically meaningful results. (5) We reiterated this process for different virtual probe paths, which were identical in shape to the path of the virtual MMS-3 but shifted away from the current sheet center along $N$ (similar to the orbits of the virtual MMS-1, 2, and 4, as is shown in Figure \ref{overview}p). The ``X'' marks on Figures \ref{mvave_errs}a, \ref{mvave_errs}b, \ref{mvave_errs}c, and \ref{mvave_errs}d indicate the errors in $L^\ast$, $M^\ast$, $N^\ast$, and $\mathcal{R}^\ast$ (respectively) that were averaged over all $10^6$ iterations of MVA-$v_e$. The two dashed curved in Figure \ref{mvave_errs}a-c are the average error plus-or-minus a standard deviation. The error in the reconnection rate, which is shown in Figure \ref{mvave_errs}d, is the difference between the average $E_{M\ast}$ ($E$ in the direction of the measured $M^\ast$ axis) and the actual average $E_M$. The red shaded area indicates the region where the virtual probe path has been moved away from the current sheet center by a distance greater than the size of the tetrahedron. While our original virtual probe path may be imperfect, any errors in the $N$ location of the virtual probe are is likely well within the white region. The data along the orbits in the red-shaded region (which are identical in shape to the orbit of virtual MMS-3, but shifted southward, away from the current sheet, by $\Delta N$) differ considerably from the observations of MMS.

First, we note that the errors in the coordinate axes due to the noise in $v_e$ are not very large. Nevertheless, the total errors in $L^\ast$, $M^\ast$, and $N^\ast$ are considerable even for the lowest-error-scenario of virtual MMS-3 ($\Delta N=0$), being approximately $5^\circ\pm2^\circ$ for $\theta_{LL\ast}$ and $\theta_{MM\ast}$ and $4^\circ\pm2^\circ$ for $\theta_{NN^\ast}$. $\theta_{LL^\ast}$ and $\theta_{MM^\ast}$ increase rapidly with the distance from the current sheet center $\Delta N$, as expected, nearly doubling from the orbit of MMS-3 ($\Delta N=0$) to the orbits of MMS-2 and 4 ($|\Delta N|=0.43$ $d_{e0}$). The error terms $\theta_{LL^\ast}$ and $\theta_{MM^\ast}$ exceed $20^\circ$ in the red shaded region at $\Delta N\approx-0.7$ $d_{e0}$. In contrast, $\theta_{NN\ast}$ is somewhat stable over the displayed range of $\Delta N$, though it begins to increase around $\Delta N\leq-0.8$ $d_{e0}$ when the noisy variance of $v_{eN}$ approaches the magnitude of the total (physical and noisy) variance of $v_{eL}$ and $v_{eM}$. For these virtual paths, the probes are far enough away from the electron current and jet layer that the variations in $v_{eL}$ and $v_{eM}$ are comparable in magnitude to those of $v_{eN}$. The small standard deviations of $\theta_{LL^\ast}$, $\theta_{MM\ast}$, and $\theta_{NN\ast}$ relative to their mean values indicate that the largest source of error is the incorrect assumption that the eigenvectors of MVA-$v_e$ are identical to the $LMN$ coordinate axes. When MVA is applied to the noiseless virtual $v_e$ data, there is no change to the average $\theta_{LL\ast}$ for small $\Delta N\geq-0.5$ $d_{e0}$, while the average values of $\theta_{MM\ast}$ and $\theta_{NN\ast}$ are both reduced by $\sim1^\circ-2^\circ$.

The error in the reconnection rate (Figure \ref{mvave_errs}d) is moderate for even the ``best-case scenario'' of $\Delta N=0$ and extreme for the ``worst-case scenario'' of $\Delta N\leq-2$ $d_{e0}$ away from the current sheet center, where $20\%\pm20\%\leq E_{M\ast}\leq80\%\pm60\%$. The noise in $v_e$ can influence $E_{M\ast}$ considerably in some cases, as the standard deviation of $E_{M\ast}$ over the $10^6$ iterations of MVA-$v_e$ is comparable to the mean. This is not unexpected, as we have already mentioned that (1) a finite projection of $M^\ast$ onto $N$ will be more likely to create errors in $E_{M\ast}$ than a finite projection of $M^\ast$ onto $L$, given that $E_N$ is typically much stronger than $E_L$ and $E_M$ and (2) the noise affects $\theta_{MM\ast}$ and $\theta_{NN\ast}$ more than $\theta_{LL\ast}$. This point is demonstrated in Figure \ref{mvave_errs}e, which shows $|90^\circ-\theta_{LM\ast}|$ (the degree to which $M\ast$ is non-orthogonal with $L$) on the horizontal axis, $|90^\circ-\theta_{NM\ast}|$ on the vertical axis, and the average value of $E_{M\ast}$ per bin as the 3rd dimension (color bar). As is evident, $E_{M\ast}$ is much larger when $M^\ast$ has a finite projection onto $N$ than it is when $M^\ast$ has the same sized projection onto $L$. {In the previous section, we estimated $\theta_{NM\ast}\approx1.3^\circ$, which corresponds to a relatively small error in $E_{M\ast}$ of $\sim$10-15\% (Figure \ref{mvave_errs}d).}

\begin{figure*}
\noindent\includegraphics[width=39pc]{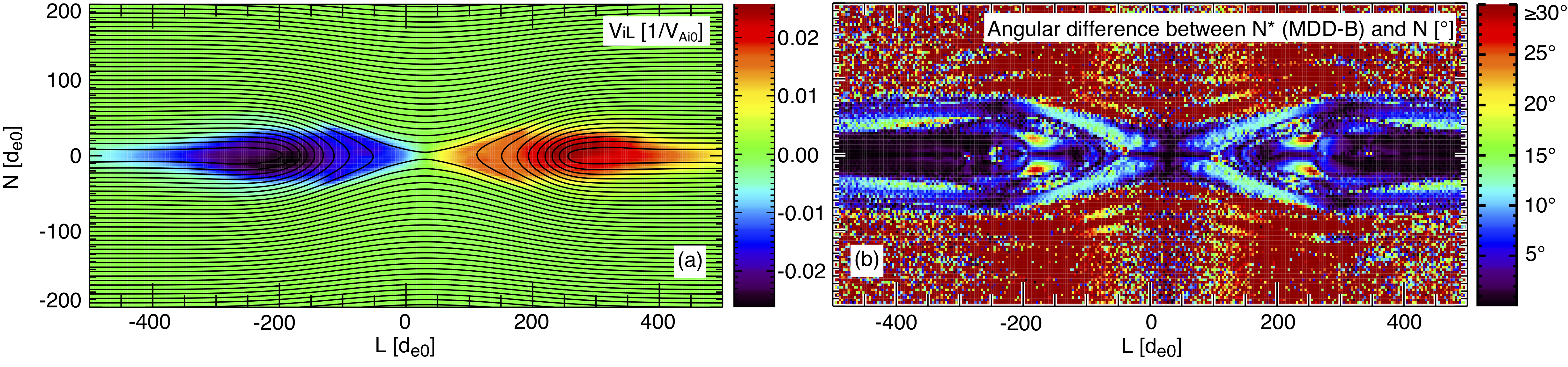}
\caption{(Right): the difference between the maximum directional derivative of $\vec{B}$ direction and the current sheet normal when MDD-$B$ is applied to every cell of the simulation near the diffusion region. (Left): for scale, the ion bulk velocity in the jet direction.}
\label{mddb2d}
\end{figure*}

Next we consider MDD-$B$. Unlike MVA-$v_e$, MDD-$B$ can be applied to every point in space without integrating or averaging over a flight path, as has been done in Figure \ref{mddb2d}b. Without accounting for MMS-like errors, MDD-$B$ is able to identify the $L$ and $N$ directions almost exactly for all points near the EDR and central current sheet. The errors in the $L^\ast$ and $N^\ast$ direction are large near the separatrices and jet braking regions, where real currents cause strong gradients in $\vec{B}$ that are not aligned with $N$. There are also large errors in the inflow region, where the noisy spatial fluctuations in $\vec{B}$ are comparable to the very small physical gradients. The $M$ direction can be identified perfectly at all points in space since it is exactly invariant in our 2-d simulation. We also find very small errors in $L^\ast$, $M^\ast$, and $N^\ast$ when MDD-$B$ is applied to the virtual tetrahedron data using the linear gradient technique (not pictured). The errors introduced by the linear gradient assumption are $\sim1^\circ-2^\circ$ for $L^\ast$ and $N^\ast$ and $\leq0.5^\circ$ for $M^\ast$. Given that the virtual tetrahedron is similar to the actual MMS tetrahedron in terms of size and regularity, these errors from non-linear gradients should be directly comparable to those that we expect from MMS if the magnetic field data were perfectly calibrated.

Given that the errors associated with the assumptions of MDD-$B$ and the linear gradient technique are small, we expect the dominant source of error (excluding possible effects from time evolution and 3-d structure) to be related to the inter-calibration of the magnetic field data rather than noise, as the errors in MDD-$B$ caused by noise were shown to be small in \cite{Denton.2012}. We make the virtual data more MMS-like by adding very small and semi-random errors to the 4-virtual-probe magnetic field data, applying MDD-$B$, then reiterating, much like we did previously for MVA-$v_e$. The errors in the virtual spin-axis ($\sim$N) and virtual-spin-plane ($\sim$L and $\sim$M) components of $\vec{B}$ are treated differently. The four-virtual-probe spin-axis errors are added as random constant offsets, which are between +0.05 nT and --0.05 nT (i.e., $\sim\pm0.0042B_b$). Smaller random and constant offsets between +0.002 nT and --0.002 nT were added to $B_L$ and $B_M$. Spin-tone-like offsets were also added to $B_L$ and $B_M$ with 90$^\circ$ phase differences. The amplitudes of the spin-tones were fixed at 0.02 nT but the differences between the phases of the virtual probes were chosen at random. In total, the absolute error assigned to any one of the virtual $\vec{B}$ measurements was no more than 0.06 nT, which is well below the 0.1 nT reported accuracy of FGM but comparable to the inter-probe differences in $\vec{B}$ observed during a quiet interval following the 2017-07-11 event.

\begin{figure*}
\noindent\includegraphics[width=24pc]{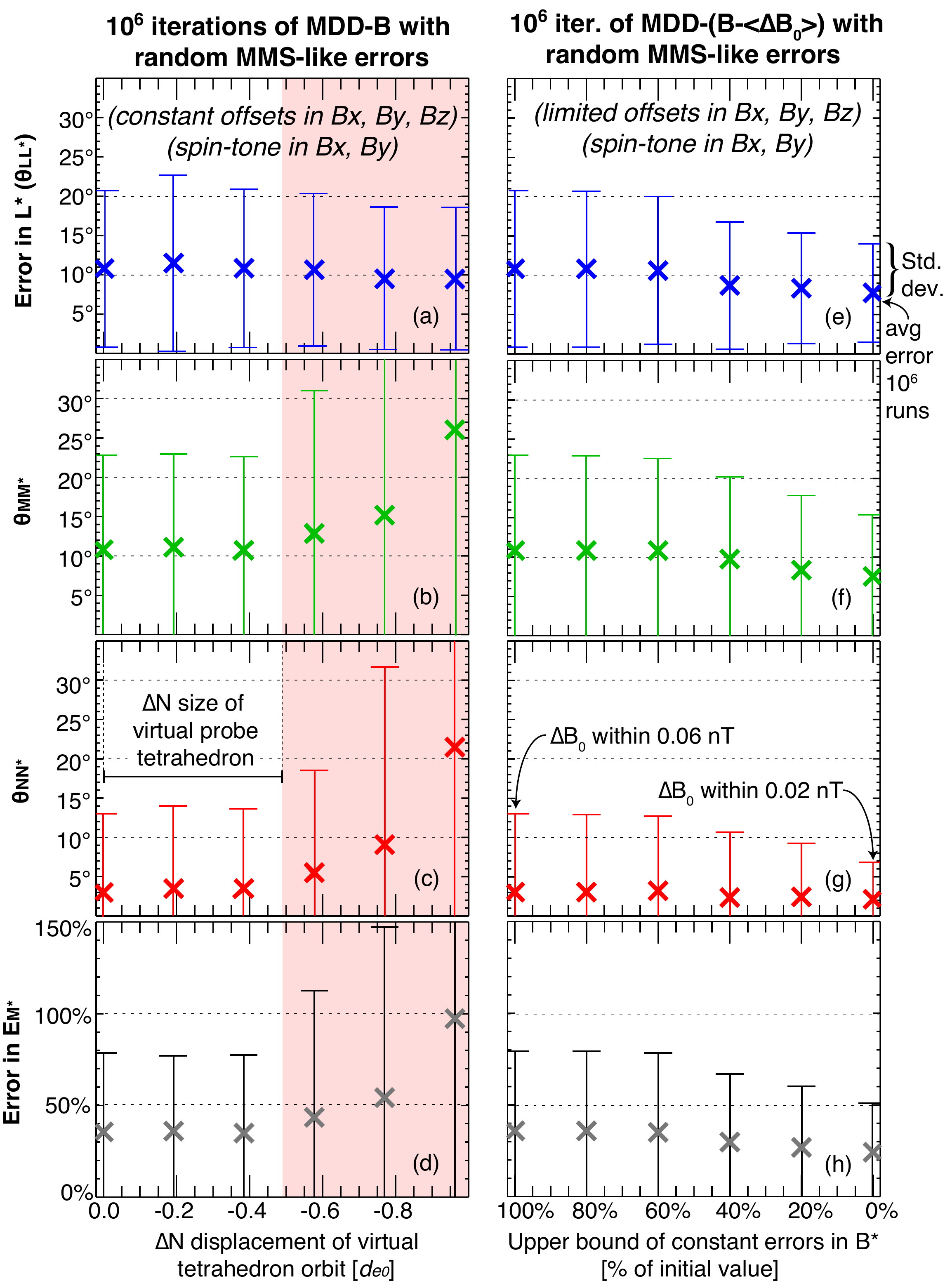}
\caption{The errors in (a, e): $L^\ast$, (b, f): $M^\ast$, (c, g): $N^\ast$, and (d, h): the reconnection rate when MDD-$B$ is applied to four-point virtual magnetic field data with added ``MMS-like'' offsets. In (a)-(d), the magnetic field data {offsets are} within 0.06 nT and the distance of the virtual tetrahedron orbit from the current sheet center is varied from the initial position determined by N18. In (e)-(h), only the virtual tetrahedron orbit from N18 is used, but the upper limits of the constant-in-time-and-space offsets, which are added to the virtual $\vec{B}$ data, are reduced from 0.06 nT (constant offsets of $\leq$0.05 nT and spin-tone offsets of $\leq$0.02 nT) to 0.02 nT (no constant offsets and spin-tone offsets of $\leq$0.02 nT).}
\label{mddb_errs}
\end{figure*}

The error terms $\theta_{LL^\ast}$, $\theta_{MM^\ast}$, $\theta_{NN^\ast}$, and $E_{M\ast}$ for MDD-$B$ are shown in Figure \ref{mddb_errs}a-d. The horizontal axes of \ref{mddb_errs}a-d are similar to the horizontal axes of Figure \ref{mvave_errs}a-d, though in \ref{mddb_errs}a-d they represent the displacement of the entire virtual tetrahedron from its initial position, rather than the displacement of the virtual MMS-3. We find that $\theta_{LL\ast}$, $\theta_{MM\ast}$, $\theta_{NN\ast}$, and $E_{M\ast}$ are quite large, even within $\Delta N\geq-0.8$ $d_{e0}$. Unlike for MVA-$v_e$, $\theta_{LL\ast}$, $\theta_{MM\ast}$, $\theta_{NN\ast}$, and $E_{M\ast}$ do not change significantly within $\Delta N\geq-0.5$ $d_{e0}$. Also unlike MVA-$v_e$, the errors in the coordinate axes and reconnection rate are almost entirely due to measurement errors. This point is clearly illustrated by the case of $\theta_{MM\ast}$, which is shown in Figure \ref{mddb_errs}b. Even though the derivative of $\vec{B}$ is set to be exactly zero in the $M$ direction, the average error in $M^\ast$ is at least $10^\circ(\pm12^\circ)$. Similar values of $\theta_{LL\ast}$ are observed, which remains more-or-less constant as a function of $\Delta N$. The average errors in the $N\ast$ direction are roughly three times smaller than the average $\theta_{LL\ast}$ and $\theta_{MM\ast}$ for small $\Delta N$, but both $\theta_{NN\ast}$ and $\theta_{MM\ast}$ begin to rapidly increase at large $\Delta N$. 

The (likely conservative) errors shown in Figure \ref{mddb_errs}d demonstrate that the errors in the reconnection rate can be considerable when $M$ is determined using the MDD-$B$ technique. This is not unexpected, as the eigenvalue of $\nabla\vec{B}(\nabla\vec{B})^\mathrm{T}$ associated with the invariant direction is zero, meaning that it is the direction that is most easily corrupted by error \cite{Denton.2010,Denton.2012}. In contrast, the normal direction, which is the direction that corresponds to the largest eigenvalue of $\nabla\vec{B}(\nabla\vec{B})^\mathrm{T}$, will be the most robust direction. This is consistent with the fact that $\theta_{NN\ast}<\theta_{LL\ast}<\theta_{MM\ast}$ was observed for orbits near the central current sheet (white area in Figure \ref{mddb_errs}a-d).

For Figures \ref{mddb_errs}e-h we consider a reduction in the constant offsets in $B_N$ but no reduction in the spin-tone offsets in $B_L$ and $B_M$. When the constant $\vec{B}$ offsets are removed the errors in the coordinate axes and reconnection rate are reduced, as is shown in Figures \ref{mddb_errs}e-h. This {is similar to how the MDD-$(B-\Delta B_0)$ coordinate system was determined} (section 3.1.4). The average errors in the coordinate axes and reconnection rate are reduced by 25-35\% when the largest possible random magnetic field errors are reduced from 0.06 nT to 0.02 nT. Compared to the background field of $B_b=12$ nT, these thresholds represent a fractional sensitivity of 0.5\% and $<0.2\%$. Compared to the ``worst-case scenario'', where the errors in the magnetic field are exactly 0.06 nT (rather than within 0.06 nT), the average values of $\theta_{LL\ast}$ and $\theta_{MM\ast}$ (=$35^\circ$) and $E_{M\ast}$ (=116\%) are reduced by nearly 80\% (not pictured). As was suggested previously, Figures \ref{mddb_errs}e-h show that the overall errors in $\theta_{LL\ast}$, $\theta_{MM\ast}$, $\theta_{NN\ast}$, and $E_{M\ast}$ are somewhat but not entirely reduced with this technique, since the removal of the spin-tone would require time-dependent calibration curves. It is not clear how precisely we were able to identify and remove the constant offsets from the MMS data before finding the {adjusted} MDD-$(B-\Delta B_0)$ coordinate system. Even for the best-case scenario, for which no constant offsets have been added, the average values of $\theta_{LL\ast}$, $\theta_{MM\ast}$, and $E_{M\ast}$ are still larger than those of MVA-$v_e$. The average values of $\theta_{NN\ast}$, conversely, are typically for MDD-$B$ and MDD-$(B-\Delta B_0)$ than they are for MVA-$v_e$. 

\section{Summary and conclusions}

We have investigated the accuracy with which we can find the $LMN$ boundary-normal coordinate system and reconnection rate $E_M$ of the MMS magnetotail electron diffusion region (EDR) event on 2017-07-11 at 22:34 UT. Overall, our results indicate that the reconnection electric field was between $2.5\leq E_M \leq4$ mV/m, which corresponds to a normalized reconnection rate of $0.14\leq \mathcal{R} \leq 0.22$ (assuming the normalization parameter is $E_b$=18.12 mV/m). We concluded that the most reliable coordinate systems are determined for this event by (1) applying MVA-$v_e$ to the probe nearest the neutral sheet, where the electron jet reversal is most pronounced, (2) applying MDD-$B$ after approximating and removing the constant (in time) offsets in the four-point measurements of $\vec{B}$, (3) applying MFR to the region near the EDR where the X-line velocity appears to be constant in time, and (4) using a hybrid approach, e.g., where MVA-$v_e$ was used to determine $L$, MDD-$B$ is used to determined $N$ perpendicular to $L$, and the third coordinate axis completes the right-handed system. However, each technique had its own sources of error, implying that one technique may not be the best for finding all coordinate axes for all events. We found that the correlation between the reconnection rates determined with these five coordinate systems was strongest for the spacecraft nearest the neutral sheet (MMS-3), likely since $E_N$ -- and therefor the projections of $E_N$ onto the imprecisely measured $M$ axes -- are reduced near the neutral sheet. {Lastly, we attempted to optimize each coordinate system by determining and removing linear correlations between $E_N$ and $E_M$. In these optimized coordinates, we found that the reconnection rate was likely $E_M$=3.2 mV/m $\pm$ 0.6 mV/m and $\mathcal{R}=0.18\pm0.035$.}

We also compared the accuracy of the MVA-$v_e$ and MDD-$B$ techniques using virtual MMS data from the EDR of a 2.5-d particle-in-cell simulation of this 2017-07-11 event, which was performed in our companion paper, \textit{Nakamura et al.} [submitted] (referred to as N18 throughout this paper). We found that the largest source of error for the MVA-$v_e$ technique was the incorrect assumption that the principle variance axes of $v_e$ were identical to the principle ($LMN$) axes of the EDR. Poor separation of the maximum and intermediate variance directions lead to moderate errors in the measured $L$ and $M$ axes, which grew rapidly as a function of the distance between the virtual probe path and the center of the current sheet. Errors in $v_e$, which are assumed to come predominantly from noise due to low counts, did not have a dramatic effect on the quality of the coordinate system and reconnection rate. When determined with MVA-$v_e$, the error in the simulated reconnection rate was moderate ($\sim$20-40\%). This error was considerably smaller when the $M$ and $N$ axes were well separated, which was most often the case. The accuracy of these techniques differed when the trajectory of the virtual probes through the EDR was altered, especially for MVA-$v_e$. 

Large errors in the $L$ and $M$ coordinate axes ($\sim10^\circ-20^\circ$) and reconnection rate ($\sim50-80\%$) were found when MDD-$B$ was applied to the virtual tetrahedron data and MMS-like errors were introduced to $\vec{B}$. These errors in $\vec{B}$ were expected to result from small errors in the inter-calibration of the magnetometers. When the constant offsets in $\vec{B}$ were not considered, and only a varying spin-tone was added, the errors in the $L$ and $M$ coordinate axes ($\sim8^\circ-15^\circ$) were somewhat reduced and the errors in the reconnection rate ($\sim25-50\%$) were dramatically reduced. This was likely due to the reduction in $\theta_{NM\ast}$, the non-orthogonality of the $N$ axis and measured $M^\ast$ axis. Unlike MVA-$v_e$, the errors in the magnetic field data were likely the only source of error for MDD-$B$, as the errors due to (1) the underlying assumption that the eigenvectors of $\nabla\vec{B}(\nabla\vec{B})^\mathrm{T}$ are equivalent to the $LMN$ coordinate axes and (2) non-linear gradients of $\vec{B}$ within the virtual tetrahedron were negligible.

Lastly, we reiterate that we have only focused on one of the sources of error in $\mathcal{R}$, which comes from the inaccurate determination of $M$. Our measurements of the normalized reconnection rate were based on $E_b$=18.12 mV/m, though different but also reasonable selections of upstream parameters (see section 2.2) could have been made such that $E_b$ was 30\% larger or smaller than our chosen value. Nominally, the accuracy of the perpendicular electric field is reported as 0.5 mV/m, which is one fifth of our measured reconnection electric field. The electric field data we have used was specially calibrated for this event (see \textit{Torbert et al} [2017]), so it is not clear whether this reported accuracy is reasonable. However, all of these sources of error coexist and compound one another in a manner that has not been accounted for in this study. Given the similarity between our measurements of the reconnection rate and those of \textit{Torbert et al} [2017] and N18, it is also possible that ours are estimates are close to the real value $\mathcal{R}$. Regardless, our results indicate that one should be cautious if comparing similar reconnection rates between two or more events, since the reconnection rate for any single event can have very large error bars, which may not be easily estimated.

\appendix

\section{Coordinate system definitions}

The coordinate systems in Table A.1 are determined using the techniques outlined in section 3. The left-most column refers to the order in which each coordinate system appeared in the enumerated list of section 3.1.6.

\begin{table*}
\centering
\caption{$LMN$ coordinate system axes in GSE.}
\begin{tabular}{| c | c || l | l | l |}
\hline
[\#] & Method & [$L_X$, $L_Y$, $L_Z$] & [$M_X$, $M_Y$, $M_Z$] & [$N_X$, $N_Y$, $N_Z$] \\
\hline
1 & GSW & [0.9986, -0.0521, 0.0052] & [0.0523,  0.9980, 0.0362] & [-0.0019, -0.0361, 0.9993] \\
\hline
2 & Modeled N.S. & [0.9986, -0.0521,   0.0052] & [0.0523,  0.9966, -0.0633] & [-0.0019,  0.0635,  0.9980] \\
\hline
3 & MVA-$B$ & [0.9935, -0.1137, -0.0107] & [0.1008,  0.9168, -0.3865] & [0.0537,  0.3829,  0.9222] \\
   & (excluding RX interval) & & & \\
\hline
4 & MVA-$B$ &  [0.9984, -0.0454,  0.0334] & [0.0562,  0.8489, -0.5256] & [-0.0045,  0.5266,  0.8501] \\
   & (only RX interval) & & & \\
\hline 
5 & MVA-$E$ &  [0.9352, -0.3495,  0.0566] & [0.3497, 0.8865, -0.3030] & [0.0557, 0.3032,  0.9513] \\
   & (long C.S. crossing) & & & \\
\hline
6 & MVA-$E$ &  [0.9750, -0.2223,  0.0017] & [0.2105,  0.9208, -0.3284] & [0.0715,  0.3205,  0.9445] \\
  & (EDR interval) & & & \\
\hline
7 & MVA-$v_i$ &  [0.9677, -0.2476, -0.0482] & [0.2477,  0.9688, -0.0038] & [0.0476, -0.0083,  0.9988] \\
\hline
8 & MVA-$v_e$ &  [0.9482, -0.2551, -0.1893] & [0.1749,  0.9168, -0.3591] & [0.2651,  0.3074,  0.9139] \\
\hline
9 & MFR &  [0.9754, -0.2131,  0.0568] & [0.2202,  0.9286, -0.2986] & [0.0109,  0.3038,  0.9527] \\
\hline
10 & MDD-$B$ &  [0.8778, 0.4194, -0.2315] & [-0.4697, 0.8485, -0.2438] & [0.0942, 0.3227,  0.9418] \\
 & (one interval) & & & \\
\hline
11 & MDD-$B$ &  [0.9451, 0.2673, -0.1866] & [-0.3139, 0.9011, -0.2990] & [0.0947, 0.3225, 0.9418] \\
 & (two intervals) & & & \\
\hline
12 & MDD-$(B-\left<\Delta B\right>_0)$ &  {[0.9858, 0.0856, -0.1443]} & {[-0.1290, 0.9367, -0.3256]} & {[0.1073, 0.3395, 0.9341]} \\
\hline
13 & Hybrid MDD-$B$ / Max $J_M$ &  {[0.971, -0.216, -0.106]} & {[0.234,  0.948, -0.215]} & {[0.054,  0.233,  0.971]} \\
 & \cite{Torbert.2018} & & & \\
\hline
14 & Hybrid MDD-$B$ / MVA-$v_e$ &  [0.9482, -0.2551, -0.1893] & [0.1818, 0.9245, -0.3350] & [0.2604, 0.2832, 0.9230] \\
\hline
\end{tabular}
\end{table*}

\acknowledgments
We would like to thank everyone who contributed to the success of the Magnetospheric Multiscale (MMS) mission. We would like to thank the Scientist in the Loop (SITL), R. E. Ergun, and the SITL on deck, M. Akhavan-Tafti, for selecting this event as one in which high-resolution MMS data is available. This work was done in collaboration with the ISSI group ``Plasma Heating and Acceleration by Collisionless Magnetic Reconnection". MMS data were obtained from the Science Data Center (SDC) at https://lasp.colorado.edu/mms/sdc/. The Space Physics Environment Data Analysis Software (SPEDAS) package was used to process the MMS data. K. J. Genestreti was supported by the Austrian FFG project 847969. R. Nakamura and T. K. M. Nakamura were supported by the Austrian FWF grant number I2016-N20. R.~E. Denton was supported by NASA grant NNX14AC38G. { For the simulation data used in this research, T.K.M. Nakamura acknowledges PRACE for awarding the access to MareNostrum at Barcelona Supercomputing Center (BSC), Spain.}

\listofchanges

\end{document}